\documentclass{jkas}
\include{graphicsx}

\def\year{2019} 
\def\month{September} 
\def\volume{00} 
\def\issue{0} 
\def\beginpage{1} 
\def\endpage{99} 
\def\received{---} 
\def\accepted{---} 
\date{Received \received ; accepted \accepted}

\title{
KMT-2018-BLG-0029Lb: A Very Low Mass-Ratio {\it Spitzer} Microlens Planet
}

\author[1,2]{Andrew Gould}
\author[3]{Yoon-Hyun Ryu}
\author[4]{Sebastiano Calchi Novati}
\author[5]{Weicheng Zang}
\author[]{\\ and \\}
\author[6]{Michael D. Albrow}
\author[3,7]{Sun-Ju Chung}
\author[8]{Cheongho Han}
\author[3]{Kyu-Ha Hwang}
\author[3]{Youn Kil Jung}
\author[3]{In-Gu Shin}
\author[9]{Yossi Shvartzvald}
\author[10]{Jennifer C. Yee}
\author[3,11]{Sang-Mok Cha}
\author[3]{Dong-Jin Kim}
\author[3]{Hyoun-Woo Kim}
\author[3,7]{Seung-Lee Kim}
\author[3,7]{Chung-Uk Lee}
\author[3]{Dong-Joo Lee}
\author[3,11]{Yongseok Lee}
\author[3,7]{Byeong-Gon Park}
\author[2]{Richard W. Pogge}
\author[]{\\ (KMTNet Collaboration) \\}
\author[12]{Charles Beichman}
\author[9]{Geoff Bryden}
\author[13]{Sean Carey}
\author[2]{B. Scott Gaudi}
\author[9]{Calen B. Henderson}
\author[14]{Wei Zhu}
\author[]{\\ ({\it Spitzer} Team) \\}
\author[15,16]{Pascal Fouqu\'e}
\author[2]{Matthew T. Penny}
\author[15]{Andreea Petric}
\author[15]{Todd Burdullis}
\author[17,18]{Shude Mao}
\author[]{\\ (CFHT Microlensing Collaboration) \\}

\affil[1]{Max-Planck-Institute for Astronomy, K\"{o}nigstuhl 17, 69117 Heidelberg, Germany}
\affil[2]{Department of Astronomy Ohio State University, 140 W.\ 18th Ave., Columbus, OH 43210, USA }
\affil[3]{Korea Astronomy and Space Science Institute, Daejon
34055, Republic of Korea}
\affil[4]{IPAC, Mail Code 100-22, Caltech, 1200 E. California Blvd., Pasadena, CA 91125, USA}
\affil[5]{Physics Department and Tsinghua Centre for
Astrophysics, Tsinghua University, Beijing 100084, China}
\affil[6]{University of Canterbury, Department of Physics and
Astronomy, Private Bag 4800, Christchurch 8020, New Zealand}
\affil[7]{University of Science and Technology, Korea (UST)
Gajeong-ro, Yuseong-gu,  Daejeon 34113, Republic of Korea}
\affil[8]{Department of Physics, Chungbuk National University,
Cheongju 28644, Republic of Korea}
\affil[9]{IPAC, Mail Code 100-22, Caltech, 1200 E. California Blvd., 
Pasadena, CA 91125, USA}
\affil[10]{ Center for Astrophysics $|$ Harvard \& Smithsonian, 60 Garden
St., Cambridge, MA 02138, USA}
\affil[11]{School of Space Research, Kyung Hee University,
Yongin, Kyeonggi 17104, Republic of Korea}
\affil[12]{NASA Exoplanet Science Institute, California Institute of Technology, Pasadena, CA 91125, USA}
\affil[13]{Spitzer, Science Center, MS 220-6, California Institute of Technology,Pasadena, CA, USA}
\affil[14]{Canadian Institute for Theoretical Astrophysics, 
University of Toronto, 60 St George Street, Toronto, ON M5S 3H8, Canada}
\affil[15]{CFHT Corporation, 65-1238 Mamalahoa Hwy, Kamuela, 
Hawaii 96743, USA}
\affil[16]{Universit\'e de Toulouse, UPS-OMP, IRAP, Toulouse, France}
\affil[17]{Department of Astronomy and Tsinghua Centre for Astrophysics, 
Tsinghua University, Beijing 100084, China}
\affil[18]{National Astronomical Observatories, Chinese Academy of Sciences, Beijing 100101, China}

\def\corrauthor{
Andrew Gould
}

\def\runningauthor{
Andrew Gould
}

\def\runningtitle{
KMT-2018-BLG-0029Lb
}

\def\keywords{
gravitational microlensing: micro; planetary systems
}

\def\abstracttext{
At $q=1.81\pm 0.20 \times 10^{-5}$,
KMT-2018-BLG-0029Lb has the lowest planet-host mass ratio $q$ of
any microlensing planet to date by more than a factor of two.  
Hence, it is the
first planet that probes below the apparent ``pile-up'' 
at $q=5$--10 $\times 10^{-5}$.  The event was observed by {\it Spitzer},
yielding a microlens-parallax $\bpi_\e$ measurement.  Combined with
a measurement of the Einstein radius $\theta_\e$ from finite-source effects
during the caustic crossings, these measurements imply
masses of the host $M_{\rm host}=1.14^{+0.10}_{-0.12}\, M_\odot$ and planet 
$M_{\rm planet} = 7.59^{+0.75}_{-0.69}\,M_\oplus$, system distance 
$D_L = 3.38^{+0.22}_{-0.26}\,\,\kpc$ and projected separation 
$a_\perp = 4.27^{+0.21}_{-0.23}\,\au$.
The blended light, which is substantially brighter than the microlensed source, is plausibly due to the lens and could be observed at high resolution immediately.
}

\newcommand{\bdv}[1]{\mbox{\boldmath$#1$}}

\def\bfq{{}}
\def\au{{\rm AU}} 
 
\def\bv{{\bf v}}

\def\kms{{\rm km}\,{\rm s}^{-1}}
\def\masyr{{\rm mas}\,{\rm yr}^{-1}}
\def\muas{{\mu\rm as}}
\def\mas{{\rm mas}}
\def\kpc{{\rm kpc}}
\def\var{{\rm var}}

\def\max{{\rm max}}
\def\rel{{\rm rel}}

\def\lim{{\rm lim}}
\def\e{{\rm E}}
\def\bpi{{\bdv\pi}}
\def\bmu{{\bdv\mu}}

\def\bgamma{{\bdv\gamma}}

\def\la{{\lesssim}}

\def\apj{{ApJ}}
\def\aj{{AJ}}
\def\apjl{{ApJL}}

\def\aap{{A\&A}}
\def\pasp{{PASP}}
\def\mnras{{MNRAS}}

\begin{document}
\jkashead 


\section{{Introduction}
\label{sec:intro}}

For most microlensing planets, the planet-host mass ratio $q$ is well
determined, but the mass of the host, which is generally too faint 
to be reliably detected, remains unknown.  Hence the planet mass
also remains unknown.  One way to carry out statistical studies
in the face of this difficulty is to focus attention on the mass ratios
themselves.  \citet{suzuki16} conducted such a study, finding
a break in the mass-ratio function at $q_{\rm br}\sim 1.7\times 10^{-4}$
based on planets detected in the MOA-II survey.  \citet{ob171434} applied
a $V/V_\max$ technique to the seven then-known microlensing planets
with well measured $q<10^{-4}$ and confirmed that the slope of the
mass-ratio function declines with decreasing mass ratio in this regime.
\citet{kb170165} considered all planets with $q<3\times 10^{-4}$ and
concluded that if the mass-ratio function is treated as a broken power law,
then the break is at $q_{\rm br}\simeq 0.56\times 10^{-4}$, with a 
change in the power-law index of $\zeta>1.6$ at $2\,\sigma$.  However,
they also noted that there were no detected microlensing planets with
$q<0.5\times 10^{-4}$ and suggested that the low end of the mass-ratio 
function might be better characterized by a ``pile-up'' around
$q\sim 0.7\times 10^{-4}$ rather than a power-law break.

In principle, one might worry that the paucity of detected 
microlensing planets for
$q\la 0.5\times 10^{-4}$ could be due to poor sensitivity at these
mass ratios, which might then be overestimated in statistical studies.
However, the detailed examination by \citet{ob171434} showed that several
planetary events would have been detected even with much lower
mass ratios.  In particular, they showed that OGLE-2017-BLG-1434Lb would have
been detected down to $q=0.018\times 10^{-4}$ and that OGLE-2005-BLG-169Lb 
would have been detected down to $q=0.063\times 10^{-4}$.  Hence,
the lack of detected planets $q\la 0.5\times 10^{-4}$ remains a puzzle.

A substantial subset of microlensing planets, albeit a minority, do
have host-mass determinations.  For most of these the mass is determined
by combining measurements of the Einstein radius $\theta_\e$ and
the microlens parallax $\bpi_\e$ \citep{gould92,gould00},
\begin{equation}
M = {\theta_\e\over\kappa\pi_\e};
\quad
\pi_\rel =\theta_\e\pi_\e;
\quad
\kappa \equiv {4G\over c^2\au}\simeq 8.1\,{\mas\over M_\odot} ,
\label{eqn:massdist}
\end{equation}
where
\begin{equation}
\theta_\e = \sqrt{\kappa M\pi_\rel};
\qquad
\bpi_\e = {\pi_\rel\over\theta_\e}\,{\bmu_\rel\over\mu_\rel},
\label{eqn:thetaepie}
\end{equation}
and $\pi_\rel$ and $\bmu_\rel$ are the lens-source relative parallax and
proper motion, respectively.  While $\theta_\e$ is routinely measured
for caustic-crossing planetary events (the great majority of those
published to date), $\bpi_\e$ usually requires significant light-curve 
distortions induced by deviations from rectilinear lens-source relative 
motion caused by Earth's annual motion.  Thus, either the event must
be unusually long or the parallax parameter $\pi_\e=\sqrt{\pi_\rel/\kappa M}$
must be unusually big.  These criteria generally bias the sample
to nearby lenses, e.g., MOA-2009-BLG-266Lb \citep{mb09266}, with lens distance
$D_L\simeq 3\,\kpc$, which was the first microlens planet with a clear parallax 
measurement\footnote{Note also the earlier case of 
OGLE-2006-BLG-109Lb,c \citep{ob06109,ob06109b}, in which the $\bpi_\e$ was
measured, but with the aid of photometric constraints.}.
In a few cases, the host mass has been measured by direct detection of
its light \citep{ob03235b,ob05169ben,mb11293b,ob05169bat}, but
see also \citet{mb08310bhat}.  This approach is
also somewhat biased toward nearby lenses, although the main issue is
that the lenses are typically much fainter than the sources, in which case one
must wait many years for the two to separate sufficiently on the plane of
the sky to make useful observations.

Space-based microlens parallaxes \citep{refsdal66,gould94,os05001}
provide a powerful alternative, which is far less biased toward nearby
lenses.  Since 2014, {\it Spitzer} has observed almost 800 microlensing
events toward the Galactic bulge 
\citep{prop2013,prop2014,prop2015a,prop2015b,prop2016} with the
principal aim of measuring the Galactic distribution of planets.
In order to construct a valid statistical sample, \citet{yee15}
established detailed protocols that govern the selection and observational
cadence of these microlensing targets.  

For 2014--2018, the overwhelming
majority of targets were provided by the Optical Gravitational Lensing
Experiment (OGLE, \citealt{ogleiv}) Early Warning System 
(EWS, \citealt{ews1,ews2}),
with approximately 6\% provided by the 
Microlensing Observations for Astrophysics
(MOA, \citealt{ob03235}) collaboration.  In June 2018, the Korea Microlensing
Telescope Network (KMTNet \citealt{kmtnet}) initiated a pilot alert program,
covering about a third of its fields \citep{alertfinder}.  In order to
maximize support for {\it Spitzer} microlensing, these fields were chosen
to be in the northern Galactic 
bulge, which is relatively disfavored by microlensing
surveys due to higher extinction, an effect that hardly impacts {\it Spitzer}
observations at $3.6\mu$m.  This pilot program contributed about 17\% of all
2018 {\it Spitzer} alerts.  
None of these events had obvious planetary signatures in the
original online pipeline reductions.  However, after the 
re-reduction of all 2018 KMT-discovered events (including those found
by the post-season completed-event algorithm, \citealt{eventfinder}), 
one of these
{\it Spitzer} alerts, KMT-2018-BLG-0029, showed a hint of an anomaly
in the light curve. This triggered tender loving care (TLC) re-reductions, 
which then revealed a clear planetary candidate.

The lens system has the lowest planet-host mass ratio 
$q=0.18\times 10^{-4}$ of any microlensing planet
found to date by more than a factor of two.

\section{{Observations}
\label{sec:obs}}

\subsection{{KMT Observations}
\label{sec:kmtobs}}

KMT-2018-BLG-0029 is at (RA,Dec) = (17:37:52.67, $-27$:59:04.92),
corresponding to $(l,b)=(-0.09,+1.95)$.  It lies in KMT field BLG14, which
is observed by KMTNet with a nominal cadence of $\Gamma = 1.0\,{\rm hr}^{-1}$
from its three sites at CTIO (KMTC), SAAO (KMTS), and SSO (KMTA) using
three identical 1.6m telescopes, each equipped with a $4\,{\rm deg}^2$ camera.
The nominal cadence is maintained for all three telescopes during the 
``{\it Spitzer} season'' (which formally began for 2018 on 
HJD$^\prime={\rm HJD}-2450000=8294.7$).  But prior to this date, the cadence
at KMTA and KMTS was at the reduced rate of $\Gamma = 0.75\,{\rm hr}^{-1}$.
The change to higher cadence fortuitously occurred just a few hours before 
the start of the KMTA observations of the anomaly.

The event was discovered on 30 May 2018 during ``live testing'' of the
alert-finder algorithm, and was not publicly released until 21 June.
However, as part of the test process, this (and all) alerts were
made available to the {\it Spitzer} team (see Section~\ref{sec:spitzobs}, 
below).

The great majority of observations were carried out in the $I$ band,
but every tenth such observation is followed by a $V$-band observation 
that is made primarily to determine source colors.
All reductions for the light curve
analysis were conducted using pySIS \citep{albrow09}, which
is a specific implementation of difference 
image analysis (DIA, \citealt{tomaney96,alard98}).

\subsection{{Spitzer Observations}
\label{sec:spitzobs}}

The event was chosen by the {\it Spitzer} microlensing team at
UT 23:21 on 19 June (JD$^\prime = 8289.47$).  The observational
cadence was specified as ``priority 1'' (observe once per 
cycle of {\it Spitzer}-microlensing time) for the first two weeks 
and ``priority 2'' thereafter (all subsequent cycles).  
Because the target lies well toward the western side
of the microlensing fields, it was one of the relatively few
events that were within the {\it Spitzer} viewing zone during the
beginning of the {\it Spitzer} season.  Therefore, it was observed
(5, 4, 2, 2) times on (1, 2, 3, 4) July, compared to roughly
one time per day for ``priority 1'' targets during the main part
of the {\it Spitzer} season.

We note that the event was chosen by the {\it Spitzer}
team about five days prior to the anomaly.  However, as mentioned
in Section~\ref{sec:intro}, the anomaly could not be discerned
from the on-line reduction in any case.  The planet KMT-2018-BLG-0029Lb
will therefore be part of the {\it Spitzer} microlensing statistical
sample \citep{yee15}.

Like almost all other planetary events from the first five years 
(2014-2018) of the {\it Spitzer} microlensing program, KMT-2018-BLG-0029 
was reobserved at baseline during the (final) 2019 season in order to test 
for systematic errors, {\bfq which were first recognized in {\it Spitzer}
microlensing data by \citet{zhu17}.  See in particular, their Figure 6.
Significant additional motivation for this decision came
from the work of \citet{kb19}, who developed a quantitative 
statistical test that they applied to the \citet{zhu17} sample and 
subsamples\footnote{\bfq  In fact, this decision was made in March 2019,
i.e., two months before the arXiv posting of \citet{kb19}.  However, the
authors extensively discussed the main ideas of their subsequent paper 
at the Microlensing Workshop in New York in January 2019.
}.
}
In the case of KMT-2018-BLG-0029, there
were 15 epochs over 3.6 days near the beginning of the bulge observing
window.  This relatively high number (compared to other archival targets)
was again due to the fact that KMT-2018-BLG-0029 lies relatively far to the
west, so that there were relatively few competing targets during the first
week of observations.

The {\it Spitzer} data were reduced using customized software
that was written for the {\it Spitzer} microlensing program
\citep{170event}.

As we discuss in Section~\ref{sec:spitzremoval}, the latter half of the
2018 {\it Spitzer} data suffer from correlated residuals.  We investigate
this in detail in the Appendix, where we identify the likely cause
of these correlated errors.  We therefore remove these data from the
main analysis and only consider them within the context of the investigation
in the Appendix.

\subsection{{SMARTS ANDICAM Observations}
\label{sec:smartsobs}}

The great majority of {\it Spitzer} events, particularly
those in regions of relatively high extinction, are targeted for
$I/H$ observations using the ANDICAM dual-mode camera \citep{depoy03} mounted
on the SMARTS 1.3m telescope at CTIO.  The purpose of these observations
is to measure the source color, which is needed both to measure
the angular radius of the source \citep{ob03262} and to facilitate a 
color-color constraint on the {\it Spitzer} source flux
\citep{yee15,170event}.  For this purpose, of order a half-dozen
observations are usually made at a range of magnifications.
Indeed, five such measurements were made of KMT-2018-BLG-0029.
Each $H$-band observation is split into five 50-second dithered exposures.  

The 2018 $H$-band observations did not extend to (or even near) baseline in part
because the event is long but mainly because of engineering problems
at the telescope late in the 2018 season.  Hence, these data cover a
range of magnification $12\la A \la 33$.  We therefore obtained six
additional $H$-band epochs very near baseline in 2019.
The $H$-band data were reduced using DoPhot \citep{dophot}.

We note that in the approximations that the magnified data uniformly
sample the magnification range $A_{\rm low} \leq A \leq A_{\rm high}$
with $n$ points and that the photometric errors are constant in flux
(generally appropriate if all the observations are below sky), the
addition of $m$ points at baseline $A_{\rm base}=1$ will improve the
precision of color measurement by a factor,
\begin{equation}
{\sigma_{\rm w/o-base}\over \sigma_{\rm with-base}}
= \sqrt{1 + K{m\over m+n}};
\quad
K\equiv 12{n-1\over n+1}\biggl({\delta A \over \Delta A}
\biggr)^2,
\label{eqn:fluximprove}
\end{equation}
where $\delta A\equiv [(A_{\rm high} + A_{\rm low})/2 - A_{\rm base}]$ and
$\Delta A\equiv (A_{\rm high} - A_{\rm low})$.  Equation~(\ref{eqn:fluximprove})
can be derived by explicit evaluation of the more general formula
$\sigma({\rm slope})=\sigma_{\rm meas}/\sqrt{n\,\var(A)}$ \citep{chi2}.
Of course, the conditions underlying Equation~(\ref{eqn:fluximprove})
will never apply exactly, but it can give a good indication of the
utility of baseline observations.  In our case $K=12(4/6)(21.5/21)^2= 8.4$,
so the predicted improvement was a factor 2.4.  The actual improvement
was a factor 2.0, mainly due to worse conditions (hence larger errors)
at baseline.

\section{{Ground-Based Light Curve Analysis}
\label{sec:anal}}

\subsection{Static Models}
\label{sec:static}

\begin{figure}
\centering
\includegraphics[width=90mm]{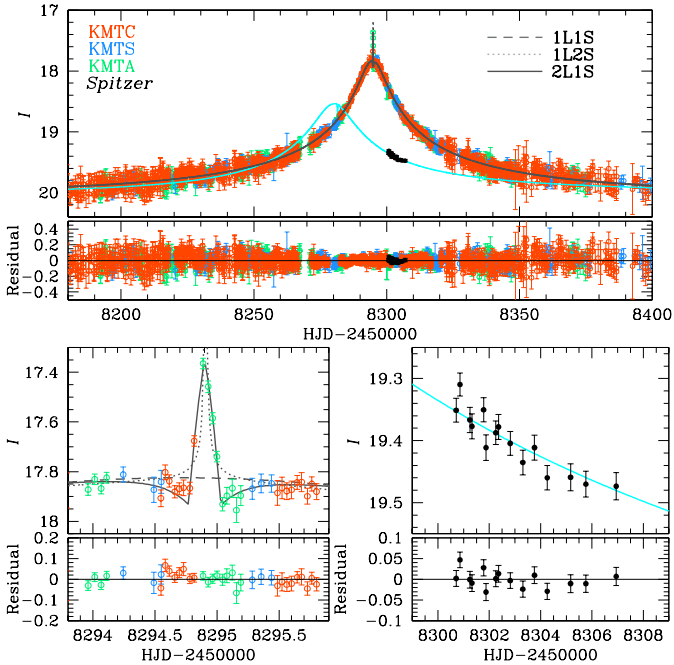}
\caption{Light curve and best fit model for KMT-2018-BLG-0029.  
The cusp crossing of the anomaly (lower-left panel) is covered by five points,
but the approaches to and from this crossing trace the overall ``dip''
that typically characterizes transverse cusp approaches.  These features
are caused by a planet with mass ratio $q=1.8\pm 0.2 \times 10^{-5}$, the
lowest of any microlensing planet to date by more than a factor two.
The {\it Spitzer} ``$L$-band'' data, which are shown in greater detail in the
lower-right panel, have been aligned 
(as usual) to the $I$-band scale by $f_{\rm display} = 
(f_L - f_{b,L})(f_{s,I}/f_{s,L}) + f_{b,I}$
(and then converted to magnitudes).  Their role in measuring the
microlens parallax $\bpi_\e$ is greatly aided by the $IHL$ color-color
relation which constrains the ratio in this expression
$(f_{s,I}/f_{s,L}) = 10^{-0.4(I-L)}$ to a few percent.  See Section~\ref{sec:IHL}.
Paczy\'nski (1L1S, dashed line) and binary-source
(1L2S, dotted line) models are clearly excluded by the data.
See Figure~\ref{fig:lcspitz} for the full {\it Spitzer} light curve, which
includes 2019 ``baseline'' data.
}
\label{fig:lc}
\end{figure}

\begin{figure}
\centering
\includegraphics[width=90mm]{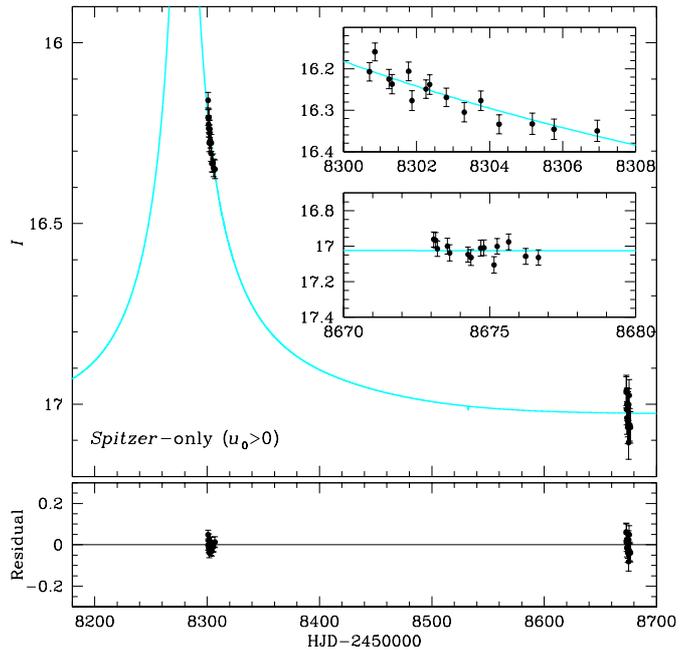}
\caption{Full Spitzer light curve including the 2018 data 
(see Figure~\ref{fig:lc}) and the 2019 ``baseline'' data.  
Only the first half of the 2018 data (covering the first six days
in time) are included in the fit and are shown here.  
See Section~\ref{sec:spitzremoval} and the Appendix for details.
}
\label{fig:lcspitz}
\end{figure}

With the exception of five ``high points'' near the peak of the event, 
the KMT light curve (Figure~\ref{fig:lc})
looks essentially like a standard single-lens single-source (1L1S)
\citet{pac86} event, which is characterized by three geometric parameters
$(t_0,u_0,t_\e)$, i.e., the time of lens-source closest approach, the
impact parameter of this approach (normalized to $\theta_\e$), and
the Einstein timescale, $t_\e=\theta_\e/\mu_\rel$.  
The five high points span just 4.4 hours, and they are flanked by
points taken about one hour before and after this interval that are
qualitatively consistent with the underlying 1L1S curve.  However,
the neighboring few hours of data on each side of the spike
actually reveal a gentle ``dip'' within which the spike erupts.
Hence, the pronounced perturbation is very short, i.e., 
of order a typical source diameter crossing time
$2t_*\equiv 2\theta_*/\mu_\rel$, where $\theta_*$ is the angular radius of
the source.  Given that the perturbation takes
place at peak, the most likely explanation is that the lens has a companion,
for which the binary-lens axis is oriented  very nearly at
$\alpha = \pm 90^\circ$ relative 
to $\bmu_\rel$.  Moreover, the source must be passing over either 
a cusp or a narrow magnification ridge that extends from a cusp.

Notwithstanding this naive line of reasoning, we conduct a systematic
search for binary-lens solutions.  We first conduct a grid search over
an $(s,q)$ grid, where $s$ is the binary separation in units of $\theta_\e$
and $q$ is the binary mass ratio.  We fit each grid point with a
seven-parameter (``standard'') model $(t_0,u_0,t_\e,s,q,\alpha,\rho)$, 
where $(s,q)$
are held fixed and the five other parameters are allowed to vary.  The
three Paczy\'nski parameters are seeded at their 1L1S values, while
$\alpha$ is seeded at six different values drawn uniformly
from the unit circle.  The last parameter, 
$\rho\equiv \theta_*/\theta_\e = t_*/t_\e$
is seeded at $\rho=(4.4\,{\rm hr})/2 t_\e\rightarrow 1\times 10^{-3}$
following the argument given above.  In addition to these non-linear
parameters there are two linear parameters for each observatory, i.e.,
the source flux $f_s$ and the blended flux $f_b$.  Hence,
the observed flux is modeled as $F(t) = f_s A(t) + f_b$, where
$A(t)$ is the time-dependent magnification at a given observatory.

\begin{table*}
\begin{center}
\caption{\textsc{Best-fit solutions for ground-only data}} 
\begin{tabular}{lccc}
\hline
\hline
\multicolumn{1}{c}{} &
\multicolumn{1}{c}{} &
\multicolumn{2}{c}{Parallax models} \\
\cline{3-4} 
\multicolumn{1}{c}{Parameters} &
\multicolumn{1}{c}{Standard} &
\multicolumn{1}{c}{$u_0>0$} &
\multicolumn{1}{c}{$u_0<0$} \\
\hline
  $\chi^2/\rm{dof}$               &1855.231/1852         &1849.908/1850         &1849.504/1850         \\
  $t_0$ $(\rm{HJD}^{\prime})$     &8294.702 $\pm$ 0.023  &8294.709 $\pm$ 0.025  &8294.704 $\pm$ 0.027  \\
  $u_0$                           &0.028 $\pm$ 0.003     &0.026 $\pm$ 0.002     &-0.027 $\pm$ 0.002    \\
  $t_{\rm E}$ $(\rm{days})$       &169.106 $\pm$ 20.595  &176.815 $\pm$ 13.742  &172.151 $\pm$ 14.743  \\
  $s$                             &1.000 $\pm$ 0.002     &0.999 $\pm$ 0.003     &1.000 $\pm$ 0.002     \\
  $q$ $(10^{-5})$                 &1.870 $\pm$ 0.243     &1.817 $\pm$ 0.267     &1.816 $\pm$ 0.215     \\
  $\alpha$ $(\rm{rad})$           &1.529 $\pm$ 0.005     &1.529 $\pm$ 0.005     &-1.529 $\pm$ 0.006    \\
  $\rho$ $(10^{-4})$              &4.603 $\pm$ 0.772     &4.414 $\pm$ 0.683     &4.577 $\pm$ 0.693     \\
  $\pi_{\rm{E},\it{N}}$           &-                     &-0.111 $\pm$ 0.084    &-0.266 $\pm$ 0.149    \\
  $\pi_{\rm{E},\it{E}}$           &-                     &0.103 $\pm$ 0.045     &0.089 $\pm$ 0.035     \\
  $\pi_{\rm{E}}$                  &-                     &0.151 $\pm$ 0.080     &0.280 $\pm$ 0.126     \\
  $\phi_\pi$                      &-                     &2.391 $\pm$ 0.570     &2.819 $\pm$ 0.673     \\
  $f_S({\rm CTIO})$               &0.029 $\pm$ 0.003     &0.028 $\pm$ 0.003     &0.029 $\pm$ 0.003     \\
  $f_B({\rm CTIO})$               &0.123 $\pm$ 0.001     &0.129 $\pm$ 0.003     &0.129 $\pm$ 0.003     \\
  $t_*$ $(\rm{days})$             &0.078 $\pm$ 0.009     &0.078 $\pm$ 0.009     &0.079 $\pm$ 0.009     \\
\hline
\end{tabular}
\tabnote{Notes, $\pi_\e\equiv\sqrt{\pi_{\e,N}^2 + \pi_{\e,E}^2}$, 
$\phi_\pi\equiv\tan^{-1}(\pi_{\e,E}/\pi_{\e,N})$, and $t_*\equiv\rho t_\e$ 
are derived quantities and are not fitted independently.  All fluxes are
on an 18th magnitude scale, e.g., $I_s= 18-2.5\,\log(f_s)$.}
\label{tab:ground}
\end{center}
\end{table*}

\begin{figure}
\centering
\includegraphics[width=90mm]{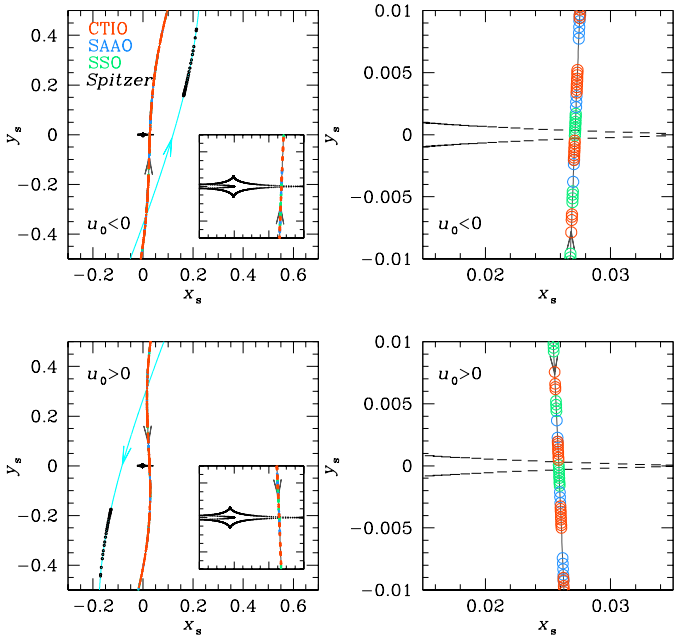}
\caption{Caustic geometries for the two parallax solutions 
($u_0>0$ and $u_0<0$).  The insets show the times of the ground-based
observations, color-coded by observatory, with the source size shown
to scale.  The right panels are zooms of these insets.
}
\label{fig:caust}
\end{figure}

This grid search yields only one local minimum, which we refine
by allowing all seven parameters to vary during the $\chi^2$
minimization.  See Figure~\ref{fig:lc} and Table~\ref{tab:ground}.  
Note that for compactness of exposition, Figure~\ref{fig:lc} shows
the {\it Spitzer} data in addition to the ground-based data.  However,
here (in Section~\ref{sec:anal}) 
we are considering results from the ground-based data alone.
See Figure~\ref{fig:lcspitz} for the full, 2018-2019 {\it Spitzer}
light curve.
As anticipated,
the binary axis is perpendicular to $\bmu_\rel$.
See Figure~\ref{fig:caust} for the caustic geometry.

\subsection{Binary Source Model}
\label{sec:binary-source}

In principle, the short-lived ``bumps'' induced on the light curve
by planets (such as the one in Figure~\ref{fig:lc}) can be mimicked
by configurations in which there are two sources (1L2S) instead of
two lenses (2L1S) \citep{gaudi98}.  Hence, unless there are obvious
caustic features, one should always check for 1L2S solutions.  In the
present case, while there are caustic features, they are less than
``completely obvious''.  

Relative to 1L1S \citep{pac86} models, the 1L2S model has four
additional parameters: the $(t_0,u_0)_2$ peak parameters of the second source,
$\rho_2$, i.e., the radius ratio of the second source to $\theta_\e$, 
and $q_{F,I}$, the $I$-band flux ratio of the second source to the first.

Figure~\ref{fig:lc} shows the best-fit 1L2S model, and Table~\ref{tab:1L2S}
shows the best-fit parameters.  For completeness, this table also
shows the best fit 1L1S model.  The 1L2S model has $\Delta\chi^2=130$
relative to the standard 2L1S model.  Moreover, it does not
qualitatively match the features of the light curve, as shown
in Figure~\ref{fig:lc}.  Therefore, we exclude 1L2S models.

\begin{table}
\begin{center}
\caption{\textsc{Best-fit solutions for 1L1S and 1L2S models}} 
\begin{tabular}{lcc}
\hline
\hline
\multicolumn{1}{c}{Parameters} &
\multicolumn{1}{c}{1L1S} &
\multicolumn{1}{c}{1L2S} \\
\hline
  $\chi^2/\rm{dof}$                   &2544.293/1856         &1985.237/1852         \\
  $t_{0,1}$ $(\rm{HJD}^{\prime})$     &8294.715 $\pm$ 0.022  &8294.639 $\pm$ 0.025  \\
  $u_{0,1}$                           &0.026 $\pm$ 0.003     &0.031 $\pm$ 0.003     \\
  $t_{\rm E}$ $(\rm{days})$           &179.591 $\pm$ 17.963  &156.531 $\pm$ 12.943  \\
  $t_{0,2}$ $(\rm{HJD}^{\prime})$     &-                     &8294.908 $\pm$ 0.002  \\
  $u_{0,2}$ $(10^{-5})$               &-                     &1.101 $\pm$ 3.348     \\
  $\rho_2$ $(10^{-4})$                &-                     &1.305 $\pm$ 0.785     \\
  $qF,I$ $(10^{-3})$                  &-                     &1.851 $\pm$ 0.187     \\
  $f_S$                               &0.028 $\pm$ 0.003     &0.032 $\pm$ 0.003     \\
  $f_B$                               &0.125 $\pm$ 0.001     &0.122 $\pm$ 0.001     \\
\hline
\end{tabular}
\label{tab:1L2S}
\end{center}
\end{table}

\subsection{Ground-Based Parallax}
\label{sec:ground-par}

Because the event is quite long, $t_\e>100\,$day, the ground-based
light curve alone is likely to put significant constraints on the
microlens parallax $\bpi_\e$.  It is important to evaluate these
constraints in order to compare them with those obtained from 
the {\it Spitzer} light curve, as a check against possible systematics
in either data set.  We therefore begin by fitting for parallax
from the ground-based light curve alone, introducing two additional
parameters $(\pi_{\e,N},\pi_{\e,E})$, i.e., the components of $\bpi_\e$
in equatorial coordinates.  

We also introduce two parameters for linearized
orbital motion $\bgamma \equiv ((ds/dt)/s,d\alpha/dt)$ because
these can be correlated with $\bpi_\e$ \citep{mb09387,ob09020}.
Here $ds/dt$ is the instantaneous rate of change of $s$, and
$d\alpha/dt$ is the instantaneous rate of change of $\alpha$, both
evaluated at $t_0$.  We expect (and then confirm) that $\bgamma$
may be relatively poorly constrained and so range to unphysical
values.  We therefore limit the search to $\beta<0.8$, where
$\beta$ is the ratio of projected kinetic to potential energy
\citep{ob05071b},
\begin{equation}
\beta\equiv \bigg|{\rm KE_\perp\over PE_\perp}\bigg|
= {\kappa M_\odot(\rm yr)^2\over 8\pi^2}{\pi_\e\over\theta_\e}\gamma^2
\biggl({s\over \pi_\e + \pi_s/\theta_\e}\biggr)^3 ,
\label{eqn:betadef}
\end{equation}
and where we adopt 
$\theta_* =0.70\,\muas$ from Section~\ref{sec:thetae-murel} (and thus,
$\theta_\e=\theta_*/\rho$) and $\pi_s=0.12$ for the source parallax.
We note that
while bound orbits strictly obey $\beta<1$, we set the limit slightly
lower because of the extreme paucity of highly eccentric planets,
and the very low probability of observing them at a phase and
orientation such that $\beta>0.8$.  We find that with $\beta$
(and thus $\bgamma$) so restricted, $\bgamma$ is neither
significantly constrained nor strongly correlated with $\bpi_\e$.
Hence, we eliminate it from the fit\footnote{Given that space-based
parallax measurements can in principle break the degeneracy
between $\bpi_\e$ and $\bgamma$ \citep{ob150768}, we again attempt to introduce
$\bgamma$ into the combined space-plus-ground fits in 
Section~\ref{sec:full-par}.
However, we again find that $\bgamma$ is neither significantly
constrained nor significantly correlated with $\bpi_\e$.  Hence,
we suppress $\bgamma$ for the combined fits as well.
}.

As usual, we check for a degenerate solution with 
$u_0\rightarrow -u_0$ \citep{smp03}, which is often called the
``ecliptic degeneracy'' because it is exact to all orders on the ecliptic
\citep{ob03238}, and which can be extended to binary and higher-order
parameters \citep{ob09020}.  Indeed, we find a nearly perfect degeneracy.
See Table~\ref{tab:ground}.

Before incorporating the {\it Spitzer}
data we must first investigate the color properties of the source.

\section{{Color-Magnitude Diagram (CMD)}
\label{sec:cmd}}

The source color and magnitude are important for two reasons.
First, they enable a measurement of $\theta_*$, and so of
$\theta_\e=\theta_*/\rho$ \citep{ob03262}.  Second, one can combine
the source color with a color-color relation to derive a constraint
on the {\it Spitzer} source flux \citep{yee15,170event}.
Table~\ref{tab:phot} lists many photometric properties of the source.

\begin{table}
\begin{center}
\caption{\textsc{Derived Photometric Properties of Source}}
\begin{tabular}{lc}
\hline
\hline
\multicolumn{1}{c}{Quantity} &
\multicolumn{1}{c}{mag} \\
\hline
$A_I$           &      $3.39$             \\
$I_{s,\rm pydia}$  &      $22.02\pm 0.08$    \\
$I_{s,\rm stand}$  &      $21.84\pm 0.12$   \\
$H_{s}$          &      $18.24\pm 0.08$   \\
$(I_{\rm pydia}-H)_s$ &   $3.78\pm 0.02$     \\
$(I-H)_{0,s}$       &    $0.87\pm 0.03$   \\
$(V-I)_{0,s}$       &    $0.78\pm 0.03$   \\
$(V-K)_{0,s}$       &    $1.71\pm 0.07$   \\
\hline
\end{tabular}
\tabnote{Note, Instrumental $I_{\rm pydia}$ is calibrated to standard $I$
from the tabulated extinction and the known position of the clump.
$H$-band data are on VVV system.} 
\label{tab:phot}
\end{center}
\end{table}

\subsection{{Source Position on the CMD}
\label{sec:source_cmd}}

The source is heavily extincted, $A_I\simeq 3.39$ 
(\citealt{gonzalez12}, where we adopt $A_I = 7\,A_K$ from a regression 
of $A_I$ from \citealt{nataf13} on $A_K$ from \citealt{gonzalez12}).
Therefore, the 
$V$-band data that are routinely taken by KMT are too noisy to
measure a reliable source color.  However, as discussed in 
Section~\ref{sec:obs}, KMT-2018-BLG-0029 (similar to most {\it Spitzer} 
targets) was observed at five epochs in $H$ band
and then was additionally observed at six epochs near baseline.

\begin{figure}
\centering
\includegraphics[width=90mm]{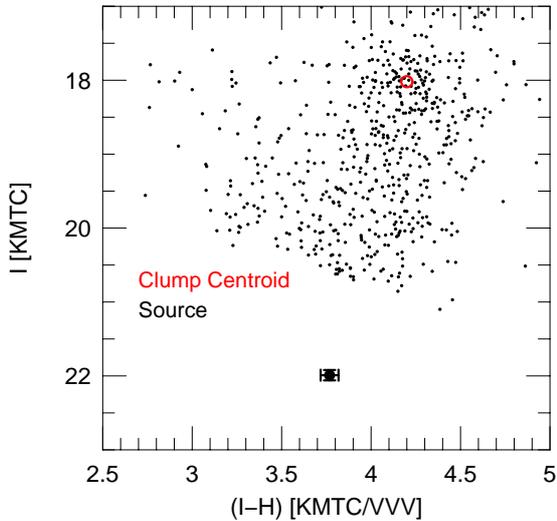}
\caption{Color-magnitude diagram (CMD) for stars within a $2^\prime$ square
centered on KMT-2018-BLG-0029.  The $I$-band data come from pyDIA reductions
of KMTC data while the $H$-band data come from the VVV catalog.  The
source $I$ magnitude (black) derives from the fit to the light curve while the
source $(I-H)$ color comes from regression of the SMARTS ANDICAM $H$-band
light curve (aligned to the VVV system) on the $I$-band light curve.
The red giant clump centroid is shown in red.
}
\label{fig:cmd}
\end{figure}

We can therefore place the source on an instrumental $(I-H,I)$ CMD
by combining these observations with the $I$-band observations from KMTC,
which is located at the same site as the SMARTS telescopes.  
To do so, we first reduce the KMTC
light curve and photometer the stars within a $2^\prime\times 2^\prime$ 
square on the same instrumental system using pyDIA.  We then 
evaluate the $(I-H)$ instrumental color by regression, finding
$(I_{\rm pyDIA} - H_{\rm ANDICAM}) = -1.035\pm 0.019$.  In order to apply
the method of \citet{ob03262} we must compare this color to that
of the red giant clump.  However, the ANDICAM data do not go deep enough
to reliably trace the clump.  We therefore align the ANDICAM
system to the VVV survey \citep{vvvcat}, finding 
$(H_{\rm ANDICAM}- H_{\rm VVV})=4.817\pm 0.005$ and therefore
$(I_{\rm pyDIA} - H_{\rm VVV}) = 3.782\pm 0.019$.   We then find
$I_{\rm pyDIA} = 22.02\pm 0.02$ by fitting the pyDIA light curve
to the best model from Section~\ref{sec:ground-par}.  We form
an $(I-H,I)$ CMD by cross-matching the KMTC-pyDIA and VVV field stars.
Figure~\ref{fig:cmd} shows the source position on this CMD.

\subsection{{$\theta_\e$ and $\mu_\rel$}
\label{sec:thetae-murel}}

We next measure the clump centroid on this CMD, finding
$[(I-H),I]_{\rm clump} = (4.20,18.02)\pm (0.02,0.04)$, which then
yields an offset from the clump of 
$\Delta[(I-H),I] =(-0.42,3.98)\pm (0.02,0.03)$.  We adopt
$[(V-I),I]_{0,\rm clump}=(1.06,14.45)$ from \citet{bensby13} and \citep{nataf13},
and use the color-color relations of \citet{bb88}, to derive
$(V-K,V)_0=(1.71,19.21)$.  That is, the source is a late G star that
is very likely on the turnoff/subgiant branch.
Applying the color/surface-brightness relation
of \citet{kervella04}, we find,




\begin{equation}
\theta_* = 0.70\pm 0.05\,\muas .
\label{eqn:thetastar}
\end{equation}
Combining Equation~(\ref{eqn:thetastar}) with $\rho$ and $t_\e$ from 
the ground-based parallax solutions in Table~\ref{tab:ground}, this implies,
\begin{equation}
\theta_\e = {\theta_*\over \rho}= 1.56\pm 0.24\,\mas;
\quad
\mu_\rel = {\theta_\e\over t_\e} =  3.3\pm 0.5\,{\mas\over \rm yr}.
\label{eqn:thetaemurel}
\end{equation}
These values strongly favor a disk lens, $D_L\la 4\,\kpc$, because otherwise
the lens would be massive (thus bright) enough to exceed the observed
blended light.  However, we
defer discussion of the nature of the lens until after incorporating the
{\it Spitzer} parallax measurement into the analysis.

\subsection{{{\it IHL} Color-Color Relation}
\label{sec:IHL}}

We match field star photometry from KMTC-pyDIA ($I$) and VVV $(H)$
(Section~\ref{sec:source_cmd}) with {\it Spitzer} $(L)$ photometry 
within the range $3.6<(I-H)<4.5$, to obtain
an $IHL$ color-color relation
\begin{equation}
I_{\rm pyDIA}-L = 1.18[(I_{\rm pyDIA}-H)_s - 3.7] + 3.32 \rightarrow 3.417\pm 0.022,
\label{eqn:IHL}
\end{equation}
where the instrumental {\it Spitzer} fluxes are converted to magnitudes
on an 18th mag system.
In order to relate Equation~(\ref{eqn:IHL}) to the pySIS magnitudes
reported in this paper (e.g., in Tables~\ref{tab:ground} and 
\ref{tab:combined}), we take account of
the offset between these two systems (measured very precisely from regression)
$I_{\rm pySIS} - I_{\rm pyDIA}= -0.120\pm 0.005$ to obtain
\begin{equation}
I_{\rm pySIS}-L = 3.297\pm 0.022,
\label{eqn:IHL2}
\end{equation}
We employ this relation when we incorporate {\it Spitzer} data in
Section~\ref{sec:spitzpar}.

\section{{Parallax Analysis Including {\it Spitzer} Data}
\label{sec:spitzpar}}

\subsection{{Removal of Second-Half-2018 {\it Spitzer} Data}
\label{sec:spitzremoval}}

As described in detail in the Appendix, we find that the second half of 
the 2018 {\it Spitzer} KMT-2018-BLG-0029 light-curve shows correlated 
residuals, and that several nearby stars display similar or related effects.
We therefore remove these epochs from the analysis.  That is, we include
only the first six days of 2018 data as well as all of the 2019 data, which
in fact were also taken during the first week (actually first 3.6 days) of the 
2019 {\it Spitzer} observing window.  We very briefly describe the essential
elements here but refer the reader to the Appendix for a thorough
discussion.

When all data are included in the analysis, there are correlated
residuals during 2018, primarily after the first week.  That is,
the light curve appears ``too bright'' during this period relative
to any model that fits the rest of the data.  There are
three bright stars within 2 {\it Spitzer} pixels, whose combined
flux is about 180 times that of the source (i.e., $f_{s,{\it Spitzer}}$).
One of these three shows a similar flux offset and another shows anomalously
larger scatter during the same period (i.e., after the first
week), but all three show essentially identical behavior between the first
week of 2018 and the first week of 2019.

All of these empirical characteristics may be explained as
due to rotation of the camera during the 
observations.  As part of normal {\it Spitzer} operations, the
camera orientation rotated at an approximately constant rate of
0.068 deg/day, i.e., by $2.5^\circ$ over the whole set of 2018 observations
but only by $0.41^\circ$ during the first six days.  The mean position
angle during this six-day period differed from the mean for 2019 by just
$0.14^\circ$, i.e., 6\% of the full rotation during 2018.  The pixel
response function (PRF, \citealt{170event}) photometry should in principle
take account of the changing pixel response as a function of camera
orientation, but if there are slight errors in the positions of the
blends due to severe crowding, then the PRF results will suffer 
accordingly.  Hence, it is plausible that the observed deviations in
both the target and blended stars, which are of order 1\% of the total
flux of the blends, are caused by this rotation.  Moreover, there can
be other effects of rotation such as different amounts of light from distant
stars falling into the grid of pixels being analyzed at each epoch
\citep{170event}. Finally, we note
that when the data are restricted to the first six days of 2018
(and first 3.6 days of 2019), the scatter about the model light curve
is consistent with the photon-noise-based photometric errors.

\subsection{{{\it Spitzer}-``Only'' Parallax}
\label{sec:spitzonly}}

\begin{table}
\begin{center}
\caption{\textsc{{\it Spitzer}-``only'' models}}
\begin{tabular}{lcc}
\hline
\hline
\multicolumn{1}{c}{Parameters} &
\multicolumn{1}{c}{$(u_0>0)$} &
\multicolumn{1}{c}{$(u_0<0)$} \\
\hline
  $\chi^2/\rm{dof}$        &26.000/26             &26.217/26          \\
  $\pi_{\rm{E},\it{N}}$    &-0.023 $\pm$ 0.037    &0.024 $\pm$ 0.037  \\
  $\pi_{\rm{E},\it{E}}$    &0.112 $\pm$ 0.008     &0.115 $\pm$ 0.007  \\
  $\pi_{\rm{E}}$           &0.115 $\pm$ 0.007     &0.117 $\pm$ 0.008  \\
  $\phi_\pi$               &1.768 $\pm$ 0.333     &1.366 $\pm$ 0.319  \\
  $f_S(Spitzer)$           &0.575 $\pm$ 0.013     &0.599 $\pm$ 0.013  \\
  $f_B(Spitzer)$           &1.871 $\pm$ 0.030     &1.845 $\pm$ 0.031  \\
\hline
\end{tabular}
\tabnote{Note, $\pi_\e$ and $\phi_\pi$ are derived quantities and are not fitted independently.  All fluxes are on an 18th magnitude scale, e.g., $L_{s,Spitzer}= 18-2.5\,\log(f_{s,Spitzer})$.}
\label{tab:spitzer}
\end{center}
\end{table}

As discussed in Section~\ref{sec:ground-par}, it is important
to compare the parallax information coming from the ground and
{\it Spitzer} separately before combining them,   in order
to test for systematics.  This remains so even though we have
located and removed an important source of systematics just above.
To trace the information coming from 
{\it Spitzer}, we first suppress the parallax information coming
from the ground-based light curve by representing it by its
seven non-parallax parameters $(t_0,u_0,t_\e,s,q,\alpha,\rho)_\oplus$ along with
the $I$-band source flux $f_{s,\oplus}$, as taken from Table~\ref{tab:ground}.
For this purpose, we use these eight non-parallax parameters taken
from the parallax solutions.  In this sense, there is some indirect
``parallax information'' coming from the ground-based fit.  However,
because we are testing for consistency, we must do this to avoid
injecting inconsistent information.  (In any case, the standard-model
and parallax-model parameters are actually quite similar.)\ \ 
We apply this procedure
separately for the two ``ecliptic degeneracy'' parallax
solutions shown in Table~\ref{tab:ground}.

\begin{figure}
\centering
\includegraphics[width=90mm]{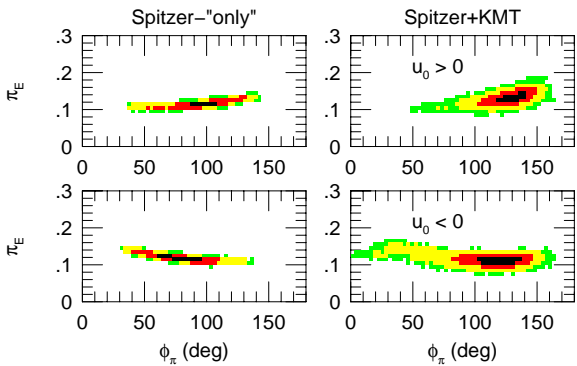}
\caption{
Likelihood contours $-2\Delta\ln L < (1,4,9)$ for 
(black, red, yellow) for the parallax vector $\bpi_\e$ in
polar coordinates.  Green indicates $-2\Delta\ln L > 9$.
Although the polar-angle $\phi_\pi$ distribution
is relatively broad for the {\it Spitzer}-``only'' fits (left panels), the
amplitude $\pi_\e$ is nearly constant because the {\it Spitzer}
observations are reasonably close to the \citet{gould12} ``cheap parallax''
limit.  See Section~\ref{sec:full-par}.  When ``one-dimensional'' parallax
information from the ground is added (right panels), the amplitude $\pi_\e$
does not qualitatively change.  See also Figures~\ref{fig:1-par} and 
\ref{fig:overlap}.
}

\label{fig:polar}
\end{figure}

\begin{figure}
\centering
\includegraphics[width=90mm]{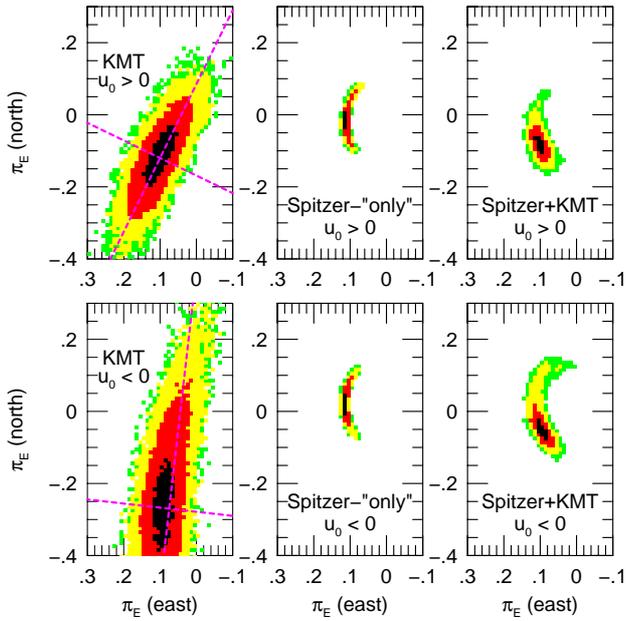}
\caption{
6-panel diagram of $(\pi_{\e,N},\pi_{\e,E})$ Cartesian contours.  
The upper panels show the $u_0>0$ solutions, while the lower panels show the
$u_0<0$ solutions.  From left to right, we display ground-only,
{\it Spitzer}-``only'', and combined parallaxes.  Black, red, and yellow
indicate relative likelihoods $-2\Delta\ln L <$ 1, 4, 9 respectively.
Green represents $-2\Delta\ln L >9$.  The ground-only data yield approximately
linear, ``one-dimensional'' constraints \citep{gmb94,smp03}.  The
{\it Spitzer}-``only'' data yield an arc opening to the west (direction
of {\it Spitzer}) because they begin post-peak and are falling rapidly
\citep{gould19}.  However, the arc is confined to an arclet of
relatively constant $\pi_\e$ amplitude (see Figure~\ref{fig:polar})
because the {\it Spitzer} observations begin when the ground data
are still highly magnified.  For at least one case ($u>0$) 
the left and center panels are consistent, implying that there
is no evidence for systematics.  Hence, the two data sets can be
combined (right).
The magenta lines in the left panels show the principal axes
defined by the $2\,\sigma$ contour.  For $u_0>0$, the contours
are nearly elliptical and the minor axis $\psi_{\rm short}=-116^\circ$
is almost perfectly aligned to the direction of the Sun at peak: $-117^\circ$,
both of which reflect ``ideal'' 1-D parallaxes.  For $u_0<0$, the
ellipse deviates from both conditions.
}
\label{fig:1-par}
\end{figure}

The left-hand panels of Figure~\ref{fig:polar} show likelihood 
contours in polar coordinates for the $u_0>0$ and $u_0<0$ solutions 
of the {\it Spitzer}-``only'' analysis.  See also Table~\ref{tab:spitzer}.
That is, $\pi_\e =|\bpi_\e|$
is the amplitude and $\phi_\pi = \tan^{-1}(\pi_{\e,E}/\pi_{\e,N})$
is the polar angle.  For both signs of $u_0$, the amplitude $\pi_\e$ is
nearly constant over a broad range of angles.  This can be understood
within the context of the argument of \citet{gould12}, which was
then empirically verified by \citet{ob161045}.  In the original argument,
a single satellite measurement at the epoch of the ground-based peak,
$t_{0,\oplus}$,
of a high-magnification event (together with a baseline measurement)
would yield an excellent measurement of $\pi_\e$ but essentially
zero information about $\phi_{\pi}$.  Because the first {\it Spitzer}
point is six days after $t_{0,\oplus}$, this condition does not strictly
hold.  However, the mathematical basis of the argument is in essence
that $u_{\rm sat}\gg u_{\oplus}$ at the time of this ``single observation''.
This is reasonably well satisfied for the first {\it Spitzer} observation.
At this time $u_\oplus\sim 0.044$.  On the other hand, 
$A(t)_{\rm spitzer}=1 + (F(t) - F_{\rm base})/F_s\rightarrow 5.0$ for the
first epoch.  
Thus\footnote{For point lenses, $u=[2((1-A^{-2})^{-1/2}-1)]^{1/2}$.}, 
$u_{Spitzer}\sim 0.203$.  If this had truly
been a single-epoch measurement, then the parallax contour would have
been an ``offset circle'' (compared to the well-centered circle
of Figure~3 of \citealt{ob161045}), with extreme parallax values
$\pi_{\e,\pm} = (\au/D_\perp)(0.203\pm 0.044)$, i.e., a factor 1.55 difference.
Here $D_\perp\sim 1.3\,\au$ 
is the projected Earth-{\it Spitzer} separation at the
measurement epoch.
However, the rest of the {\it Spitzer} light curve then restricts this
circle to an arc.  See Figures~1 and 2 of \citet{gould19}, which
also illustrate how the two {\it Spitzer}-``only'' solutions
(for a given sign of $u_0\equiv u_{0,\oplus}$) merge.
Figure~\ref{fig:1-par} shows the $\bpi_\e$ contours in Cartesian coordinates
for the six cases.  Here we focus attention on four of these cases,
(ground-only, {\it Spitzer}-``only'')$\times (u_0<0,\ u_0>0)$.

\begin{figure}
\centering
\includegraphics[width=90mm]{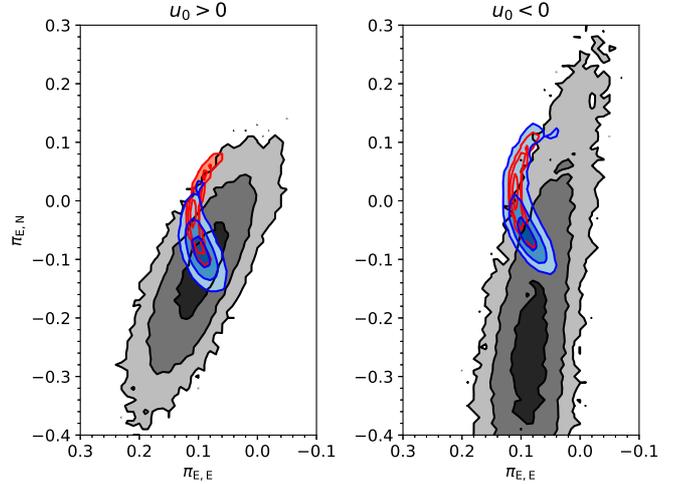}
\caption{
Overlap of three sets of contours shown in Figure~\ref{fig:1-par}
for each of the two parallax solutions.  This makes it easier to see
that for $u_0>0$ the ground-only and {\it Spitzer}-``only'' solutions
are consistent, showing that there is no evidence for systematics.
Then, the fact that these solutions show some tension for $u_0<0$ implies
that this solution is somewhat disfavored.
}
\label{fig:overlap}
\end{figure}

These show that the ground-only and {\it Spitzer}-``only'' 
parallax contours are consistent for the $u>0$ case and
marginally inconsistent for the $u<0$ case.  The levels of consistency
can be more precisely gauged from Figure~\ref{fig:overlap}, which
shows overlapping contours.  Because one of these two cases
is consistent, there is no evidence for systematics in either
data set.  That is, only one of the two cases can be physically
correct, so only if both were inconsistent would the comparison
provide evidence of systematics.

\subsection{{Full Parallax Models}
\label{sec:full-par}}

\begin{table}
\begin{center}
\caption{\textsc{Best-fit solutions for ground+{\it Spitzer} data}}
\begin{tabular}{lcc}
\hline
\hline
\multicolumn{1}{c}{} &
\multicolumn{2}{c}{Parallax models} \\
\cline{2-3} 
\multicolumn{1}{c}{Parameters} &
\multicolumn{1}{c}{$u_0>0$} &
\multicolumn{1}{c}{$u_0<0$} \\
\hline
  $\chi^2/\rm{dof}$               &1877.274/1878         &1881.580/1878         \\
  $t_0$ $(\rm{HJD}^{\prime})$     &8294.716 $\pm$ 0.025  &8294.727 $\pm$ 0.025  \\
  $u_0$                           &0.027 $\pm$ 0.003     &-0.027 $\pm$ 0.003    \\
  $t_{\rm E}$ $(\rm{days})$       &173.950 $\pm$ 15.754  &176.564 $\pm$ 16.346  \\
  $s$                             &1.000 $\pm$ 0.002     &1.000 $\pm$ 0.002     \\
  $q$ $(10^{-5})$                 &1.829 $\pm$ 0.217     &1.758 $\pm$ 0.222     \\
  $\alpha$ $(\rm{rad})$           &1.531 $\pm$ 0.005     &-1.534 $\pm$ 0.005    \\
  $\rho$ $(10^{-4})$              &4.472 $\pm$ 0.692     &4.398 $\pm$ 0.708     \\
  $\pi_{\rm{E},\it{N}}$           &-0.086 $\pm$ 0.028    &-0.054 $\pm$ 0.042    \\
  $\pi_{\rm{E},\it{E}}$           &0.100 $\pm$ 0.013     &0.093 $\pm$ 0.016     \\
  $\pi_{\rm{E}}$                  &0.132 $\pm$ 0.013     &0.107 $\pm$ 0.011     \\
  $\phi_\pi$                      &2.281 $\pm$ 0.217     &2.092 $\pm$ 0.394     \\
  $f_S({\rm CTIO})$               &0.028 $\pm$ 0.003     &0.028 $\pm$ 0.003     \\
  $f_B({\rm CTIO})$               &0.128 $\pm$ 0.001     &0.128 $\pm$ 0.001     \\
  $f_S(Spitzer)$                  &0.584 $\pm$ 0.056     &0.580 $\pm$ 0.059     \\
  $f_B(Spitzer)$                  &1.865 $\pm$ 0.054     &1.866 $\pm$ 0.056     \\
  $t_*$ $(\rm{days})$             &0.078 $\pm$ 0.009     &0.078 $\pm$ 0.009     \\
\hline
\end{tabular}
\tabnote{Note, $\pi_\e$, $\phi_\pi$, and $t_*$ are derived quantities and are not fitted independently.  All fluxes are
on an 18th magnitude scale, e.g., $I_s= 18-2.5\,\log(f_s)$.} 
\label{tab:combined}
\end{center}
\end{table}

We therefore proceed to analyze the ground- and space-based data together.  
The resulting microlens parameters for the two cases 
($u_{0,\oplus}<0$ and $u_{0,\oplus}>0$) are shown in Table~\ref{tab:combined}.  
The parallax
contours are shown in the right-hand panels of Figures~\ref{fig:polar} and
\ref{fig:1-par} and also superposed on the ground-only and 
{\it Spitzer}-``only'' contours in Figure~\ref{fig:overlap}.

The first point to note is that while the $\chi^2$ values of the
two $\pm u_{0,\oplus}$ topologies are nearly identical for the ground-only and 
{\it Spitzer}-``only'' solutions, the combined solution favors $u_0>0$
by $\Delta\chi^2 = 4.3$.  This reflects the marginal inconsistency 
for the $u_0<0$ case that we identified in Section~\ref{sec:spitzonly}.  See 
Figure~\ref{fig:overlap}.

The next point is that the effect of the ground-based parallax ellipse
(left panels of Figure~\ref{fig:1-par}) is essentially to preferentially 
select a subset of the {\it Spitzer}-``only'' arc (middle panels).  This
is especially true of the $u_0>0$ solution, which we focus on first.
The long axis of the ground-only ellipse 
(evaluated by the $\Delta\chi^2=4$ contour)
is aligned at an angle $\psi_{\rm long}\simeq -26^\circ$ north through east,
implying that the short axis is oriented at $\psi_{\rm short}\simeq -116^\circ$.
This is close to the projected position of the Sun at $t_{0,\oplus}$,
$\psi_\odot = -117^\circ$, which means that the main ground-based parallax
information is coming from Earth's instantaneous acceleration near
the peak of the event.  This is somewhat surprising because this instantaneous
acceleration is rather weak ($\sim 17\%$ of its maximum value) due to the
fact that the event is nearly at opposition.  However, it confirms that
despite the large value of $t_\e\sim 175\,$days, it is primarily the
highly magnified region near the peak, where the fractional photometry errors
are smaller, that contributes substantial parallax information.
The measurement of the component of parallax along this $\psi_{\rm short}$
direction ($\pi_{\e,\parallel}$) not only has smaller statistical errors than
$\pi_{\e,\perp}$
(as illustrated by the ellipse), but is also less subject to systematic
errors because it is much less dependent on long term photometric
stability over the season.  From inspection of the left panel of
Figure~\ref{fig:overlap}, it is clear that the intersection of the
ground-only and {\it Spitzer}-``only'' contours is unique and would
remain essentially the same even if the ground-only contours were
displaced along the long axis.

The situation is less satisfying for the $u_{0,\oplus}<0$ solution
in several respects.  These must be evaluated within the context
that, overall, this solution is somewhat disfavored by the marginal
inconsistency between the ground-only and {\it Spitzer}-``only'' 
solutions discussed in Section~\ref{sec:spitzonly}.  First, the
error ellipse is oriented at $\psi_{\rm short} \simeq -97^\circ$, which
is $20^\circ$ away from the projected position of the Sun at $t_{0,\oplus}$.
This implies that the dominant parallax information is coming
from after peak rather than symmetrically around peak, which already
indicates that it is less robust and more subject to long-timescale
systematics.  Related to this, the uncertainties 
in the $\psi_{\rm long}$ direction
are larger.  Hence, we should consider how the solution would
change for the case that systematics have shifted the ground-only error ellipse 
along the long axis by a few sigma.  From inspection of the right
panel of Figure~\ref{fig:overlap}, this would tend to create
a second, rather weak, minimum near $(\pi_{\e,N},\pi_{\e,E})\simeq (+0.16,+0.04)$.
However, even under this hypothesis, this new minimum would suffer even
stronger inconsistency between ground-only and {\it Spitzer}-``only''
solutions than the current minimum.

We conclude that the $u_0<0$ solution is disfavored, and even
if it is nevertheless correct, its parallax is most likely
given by the displayed minimum rather than a secondary minimum that
would be created if the ground-based contours were pushed a few
sigma to the north.
Moreover, the parallax amplitude $\pi_\e=|\bpi_\e|$
is actually similar for the two
minima (see lower panels of Figure~\ref{fig:polar}), and it is only
$\pi_\e$ that enters the mass and distance determinations.  We conclude
that the physical parameter estimates, which we give in Section~\ref{sec:phys},
are robust against the typical systematic errors that are described above.

Nevertheless, we will conduct an additional test in the space of physical
(as opposed to microlensing) parameters.  However, we defer this test
until after we derive the physical parameters from the microlensing
parameters in Table~\ref{tab:combined}.

\section{{Physical Parameters}
\label{sec:phys}}

\begin{table}
\begin{center}
\caption{\textsc{Physical parameters for Ground+Spitzer models}}
\begin{tabular}{lcc}
\hline
\hline
\multicolumn{1}{c}{Quantity} &
\multicolumn{1}{c}{$u_0>0$} &
\multicolumn{1}{c}{$u_0<0$} \\
\hline
  $M_{\rm host}$ $[M_\odot]$     &$1.36_{-0.22}^{+0.25}$  &$1.57_{-0.26}^{+0.28}$  \\
  $M_{\rm planet}$ $[M_\oplus]$ &$8.44_{-1.02}^{+1.19}$  &$9.85_{-1.15}^{+1.28}$  \\
  $a_{\bot}$ [au]               &$4.63_{-0.38}^{+0.41}$  &$5.06_{-0.42}^{+0.42}$  \\
  ${\it D_L}$ [kpc]             &$3.21_{-0.23}^{+0.28}$  &$3.52_{-0.22}^{+0.28}$  \\
  $\mu_{{\rm hel},N}$ [mas/yr]  &$-1.92_{-0.50}^{+0.57}$ &$-1.39_{-0.76}^{+1.07}$ \\
  $\mu_{{\rm hel},E}$ [mas/yr]  &$3.51_{-0.52}^{+0.53}$  &$3.62_{-0.56}^{+0.57}$  \\
  $v_{{\rm L,LSR},l}$ [km/s]    &$-71_{-52}^{+52}$         &$-71_{-59}^{+59}$       \\
  $v_{{\rm L,LSR},b}$ [km/s]    &$-54_{-41}^{+41}$         &$-57_{-45}^{+45}$         \\
\hline
\end{tabular}
\label{tab:phys}
\end{center}
\end{table}

We evaluate the physical parameters of the system by directly
calculating their values for each element of the Monte Carlo Markov chain
(MCMC).  In particular, for each element, we evaluate
$\theta_* = \theta_{*,0}(1+\epsilon_*)$, where
$\theta_{*,0} = 0.70\,\muas (f_{s,\rm pySIS}/0.028)^{1/2}$
and $\epsilon_*=4\%$ is
treated as a random variation.  However, we note that the largest
source of uncertainty in $\theta_\e$ is the $\sim 15\%$
error in $\rho$.  These physical parameters are reported in 
Table~\ref{tab:phys}.
For our analysis, we adopt a source distance $D_S=R_0=8.2\,\kpc$,
and source motions in the heliocentric frame drawn from a 
distribution derived from {\it Gaia} data\footnote{Because the actual line
of sight $(l,b)=(-0.09,+1.95)$ 
is heavily extincted, we evaluate the {\it Gaia} proper-motion ellipse
at the symmetric position $(l,b)=(-0.09,-1.95)$.  We consider stars
within a $2^\prime$ square of this position and restrict attention
to Bulge giants defined by $G<18$ and $B_p - R_p>2.25$.  We eliminate
four outliers and make our evaluation based on the remaining 226 stars,
the majority of which are clump giants.},
$\bmu_s(l,b) = (-5.7,0.0)\,\masyr$, $\sigma(\bmu_s)=(3.4,2.7)\,\masyr$.

We note that while the central values for the lens velocity in
the frame of the local standard of rest (LSR) are large, they
are consistent within their $1\,\sigma$ errors with typical values for 
disk objects.  These large errors are completely dominated by the
uncertainty in the source proper motion, which propagates to errors
in $\bv_{l,LSR}$ of $D_L\sigma(\bmu_s)= (48,38)\,\kms\,(D_L/3\,\kpc)$. 
These are then added in quadrature to the much smaller
terms from other sources of error.

We next test whether the lens mass and distance estimates shown
in Table~\ref{tab:phys} are consistent with limits on lens light
in baseline images.  For this purpose, we take $r$ and $i$ images
using the 3.6m Canada-France-Hawaii Telescope (CFHT) at Mauna Kea, Hawaii,
which are both deeper and at higher resolution than the KMT image.
We align the two systems photometrically and find 
$I_{\rm base,pyDIA}= 20.085\pm 0.044$, which implies blended flux
(in these higher resolution images) of $I_{b,\rm pyDIA}=20.29\pm 0.07$.
We note that the error bar, which is derived from the photometry routine,
implicitly assumes a smooth background, which is not the case
for bulge fields with their high surface-density of background stars.
We ignore this issue for the moment but treat it in detail in 
Section~\ref{sec:mottled}.
We then compare the position of the clump $I_{\rm cl,pyDIA}=18.02$
to that expected from standard photometry \citep{nataf13} and
the estimated extinction $A_I=3.39$, i.e., $I_{\rm cl,stand}=17.84$
to derive a calibration offset $\Delta I=-0.18\pm 0.09$.  This
yields $I_{b,\rm stand} = 20.11\pm 0.12$.

In asking whether the upper limits on lens flux implied by this blended 
light are consistent with the physical values in Table~\ref{tab:phys},
we should be somewhat conservative and assume that the lens lies
behind the full column of dust seen toward the bulge, $A_{I,l}=3.39$.
Then, $I_{0,b} = 16.72\pm 0.12$, and hence (incorporating the $1\,\sigma$
range of distances for the $(u_0>0)$ solution), the corresponding absolute
magnitude range is $M_{I,l} = 4.19\pm 0.21$.  This
range is consistent at the $1\,\sigma$ level with the expectations 
for the  $M_{\rm host}=1.36^{+0.25}_{-0.22}$ host reported for the $(u_0>0)$ 
solution.

We conclude that the blended light is a good candidate for the light
expected from the lens. However, given the faintness of the source
and the difficulties of seeing-limited observations (even with very good
seeing), we refrain from concluding that we have in fact detected
the lens.

Nevertheless, we note that, the corresponding calculation for 
the $u_0<0$ solution
leads to mild $(\sim 1.5\,\sigma)$ tension, rather than simple 
consistency.
When combined with the earlier indications of marginal inconsistency,
we consider that overall the $u_0<0$ solution is disfavored.

\subsection{{Baseline-flux Error Due to ``Mottled Background''}
\label{sec:mottled}}

The point-spread-function (PSF) fitting routine used to derive the
flux and error of the ``baseline object'' implicitly assumes that
that this (and all detected) sources are sitting on top of a uniform
background.  It measures this background from neighboring regions
that are ``without stars'' and then subtracts this measured background
from the tapered aperture at the positions of the sources.  The lens,
the unmagnified source, as well as possible companions to either 
(which are therefore associated with the event) contribute to the
resulting ``baseline object'' light, and of course other ambient
sources that are not associated may contribute as well.  Because of
this possibility, the blended light (baseline light with source light
subtracted) can only be regarded as an upper limit on the lens
light, unless addtional measurements and/or arguments are brought to
bear.  

However, it is also possible that the entire ``mottled background''
of ambient (unrelated) stars can actually {\it reduce} the measured
baseline flux below the sum of the unmagnified source flux plus lens
flux if there is a ``hole'' in this background at the location of the
event.  This effect was first noted by \citet{mb0337} in order
to explain so-called ``negative blending''.  But it is also important
to consider this effect in the context of upper limits on lens light.

We model the distribution of background stars using the \citet{holtzman98}
$I$-band luminosity function (HLF), which is based on 
{\it Hubble Space Telescope (HST)} images toward Baade's Window (BW).  We
then increase the normalization of the HLF by a factor 2.42 because 
the surface density of bulge stars is much higher at the lens location,
$(l,b)=(-0.09,+1.95)$, than at BW.  We evaluate this normalization
factor from the ratio of the surface density of clump giants at the 
event location reflected
through the Galactic plane, $(l,-b)=(-0.09,-1.95)$, to the one at BW
(\citealt{nataf13}, D.\ Nataf 2019, private communication.)

Next, we restrict consideration to background stars more than 0.7 mag 
fainter than the ``baseline object'', i.e., $I>20.81$. Stars that are
brighter than this limit are detected by the PSF photometry program
and so do not contribute to the program's ``background light'' parameter.
Of course, brighter stars may contribute ``baseline object'' flux,
but this effect is already accounted for in the naive treatment.
Next we add $3.39 + 14.54= 17.93$ to the absolute magnitudes in the HLF
to take account of extinction and mean distance modulus.  Hence our
threshold corresponds to $M_I = 20.81 - 17.93 = 2.88$ on the HLF.  Note
that the surface density of stars at this threshold (even after multiplying
by 2.42) is only $N\sim 0.27\,{\rm arcsec}^{-2}$, or about 0.4 stars per
$\pi{\rm FWHM}^2$ seeing disk, where ${\rm FWHM}=0.7^{\prime\prime}$ is the
CFHT full width at half maximum.  That is, in this case, the threshold
is set at the detection limit rather than confusion limit.  In more
typical fields, with $A_I\lesssim 1.5$, the opposite would typically be
the case.

We then created 10,000 random realizations of the background star distribution,
and measure the excess or deficit of flux addributed to the ``baseline object''
due to this mottled background.  In order to give physical intuition to
these results, we add this excess/deficit flux to a fiducial $I=20.11$
star and ask how its magnitude changes due to this effect.  We find
at ``$1\,\sigma$'' (16th, 50th, 84th percentiles)
$\delta I = -0.04_{-0.15}^{+0.27}$ and at 
``$2\,\sigma$'' (2.5th, 50th, 97.5th percentiles)
$\delta I = -0.04_{-0.23}^{+0.76}$.

In the current context, our principal concern is the impact of these
additional uncertainties on the upper limit on lens light.  We see from the
above calculation that at the $1\,\sigma$ level, the lens could be
$-0.04-0.15=-0.19$ mag brighter than than the apparent blend flux due the
effect of a ``hole'' in the ``mottled background''.  This compares to the
$\pm 0.21$ mag error in the flux due to all factors in the comparison
of the lens to the blended flux, except for the lens mass (and chemical
composition).  Previously, we judged that the predicted lens light was
consistent with the blended light for $(u_0>0)$ solution.  Of course, increasing
these error bars does not alter that consistency.

For the $(u_0<0)$ solutions we previously judged that there 
was $1.5\,\sigma$ tension
because at the best estimates for the mass ($M\sim 1.31$--$1.85\,M_\odot$), 
the lens would be substantially brighter than the blended light. 
The additional uncertainty from the mottled-background effect raises
the $1\,\sigma$ range on the flux limit from 0.21 mag to 0.31,
which softens the inferred mass limit by just 3\%.
Hence, these larger errors do not qualitatively alter our previous
assessment of ``mild tension'' from the flux limit.

For reference, we note that for a more typical field, with $A_I=1.4$
(rather than 3.39) and a surface density 1.7 times that of BW (rather
than 2.42), we find that the $1\,\sigma$ error range would be
substantially more compact, $\delta I = -0.02_{-0.08}^{+0.12}$ 
(rather than $\delta I = -0.04_{-0.15}^{+0.27}$).



\begin{table}
\begin{center}
\caption{\textsc Physical parameters including flux constraint} 
\begin{tabular}{lcc}
\hline
\hline
\multicolumn{1}{c}{Quantity} &
\multicolumn{1}{c}{$u_0>0$} &
\multicolumn{1}{c}{$u_0<0$} \\
\hline
  $M_{\rm host}$ $[M_\odot]$   &$1.14_{-0.12}^{+0.10}$  &$1.25_{-0.12}^{+0.09}$  \\
  $M_{\rm planet}$ $[M_\oplus]$ &$7.59_{-0.69}^{+0.75}$  &$8.69_{-0.81}^{+0.78}$  \\
  $a_{\bot}$ [au]            &$4.27_{-0.23}^{+0.21}$  &$4.54_{-0.22}^{+0.16}$  \\
  ${\it D_L}$ [kpc]         &$3.38_{-0.26}^{+0.22}$  &$3.76_{-0.24}^{+0.18}$  \\
  $\mu_{{\rm hel},N}$ [mas/yr] &$-1.87_{-0.44}^{+0.50}$ &$-1.17_{-0.79}^{+1.57}$ \\
  $\mu_{{\rm hel},E}$ [mas/yr  &$3.15_{-0.39}^{+0.44}$  &$3.24_{-0.46}^{+0.44}$  \\
  $v_{{\rm L,LSR},l}$ [km/s]    &$-77_{-55}^{+55}$     &$-77_{-62}^{+67}$       \\
  $v_{{\rm L,LSR},b}$ [km/s]    &$-51_{-43}^{+43}$       &$-53_{-48}^{+49}$        \\
\hline
\end{tabular}
\label{tab:phys_flux}
\end{center}
\end{table}

\subsection{{Physical Parameter Estimates Including Flux Limit}
\label{sec:flux}}

As noted in the previous two subsections, the range of physical
parameters derived directly from the microlensing (and CMD) parameters
is consistent with the upper limit on lens light at the $1\,\sigma$
level (at least for the $u_0>0$ solution).  Nevertheless, a significant
fraction of this $1\,\sigma$ range (as well as all masses above $1\,\sigma$)
are inconsistent.  Hence, to obtain physical-parameter estimates that
reflect all available information, we should impose a flux constraint
by censoring those realizations of the MCMC that violate this constraint.
To do so, we eliminate MCMC elements that fail the condition
$D_L > 2.31\,\kpc\, (M/M_\odot)^{3/2}$, which would correspond to 
$M_{I,L} > 4.90 -7.5\,\log(M/M_\odot)$ under the assumption that the
blended light were exactly $I_{0,b}=16.72$.  The zero point of this
relation is set 0.5 mag higher than the zero-age main-sequence of the
sun ($M_I\sim 4.4$) to take account of the 0.3 mag error in $I_{0,b}$
as well as the unknown metalicity of the lens.  The slope of the
relation approximates the $I$-band luminsotiy as $\propto M^3$ over
the fairly narrow mass range where it is relevant.  That is, this
flux constraint is meant to be mildly conservative because we are seeking
the best estimates for the physical parameters rather than trying
to place very conservative limits on some part of parameter space.  The
results are given in
Table~\ref{tab:phys_flux}.  We adopt the $u_0>0$ solution from this
Table for our final estimates of the physical parameters.  
We note that the $u_0<0$ solution generally overlaps these
values at the $1\,\sigma$ level.  Hence, because this solution
is formally disfavored by a factor $>10$ due to higher $\chi^2$
and more MCMC realizations excluded by the flux condition, the final
results would barefly differ if we had adopted a weighted average
(e.g., $<0.01 M_\odot$ for the case of $M_{\rm host}$).

\section{{Bayesian Test}
\label{sec:bayesian}}

\begin{figure}
\centering
\includegraphics[width=90mm]{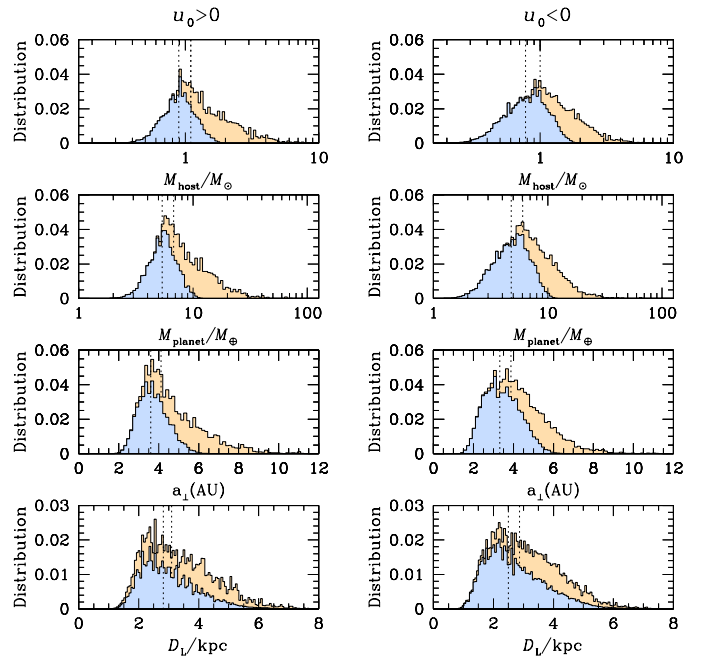}
\caption{Bayesian posteriors for four physical parameters
$(M_{\rm host},M_{\rm planet},a_\perp,D_L)$ obtained by applying constraints
from the ground-only microlensing fit (Table~\ref{tab:ground})
to simulated events from a Galactic model (yellow histograms).
The blue histograms show the results of applying the lens-flux 
constraint: $D_L > 2.31\,\kpc\,(M/M_\odot)^{1.5}$
based on limits on the lens light from the blend.  The
$u_{0,\oplus}>0$ (left) and $u_{0,\oplus}<0$ (right) solutions
are qualitatively similar, although the latter are generally
broader, both before (yellow) and after (blue) the flux
constraint is applied (blue).
}
\label{fig:bayes}
\end{figure}

Because we have measured both the microlens  parallax $\bpi_\e$
and the Einstein radius $\theta_\e$ reasonably precisely, our
main orientation has been to estimate the physical parameters
using the microlensing (and CMD) parameters alone, supplemented
by the flux constraint but without Galactic priors.  However, it is of
some interest to ask how the event would have been analyzed in the
absence of {\it Spitzer} data.

We therefore next conduct a Bayesian analysis using only the
ground-based data, i.e., ignoring the {\it Spitzer}
data.  We generally follow the procedures of \citet{ob171522}.  We represent the
outcome of the microlensing light-curve analysis by Gaussian errors
for $t_\e$ (using Table~\ref{tab:ground}) and $\theta_\e =1.56\pm 0.24\,\mas$
according to Equation~(\ref{eqn:thetaemurel}).  We represent the constraint
on $\bpi_\e$ as a 2-D Gaussian derived from the left panels of 
Figure~\ref{fig:1-par}.  Then we weight model Galactic events (as per
\citealt{ob171522}) according to these Gaussians.  
The results are shown as yellow contours in Figure~\ref{fig:bayes}.  
The resulting profiles are relatively broad, and they
peak near the results shown in Table~\ref{tab:phys} derived from the 
ground+{\it Spitzer} analysis.  For example, the median host mass for
$(u_0>0)$ is $1.1\,M_\odot$ compared to $M_{\rm host}=1.36^{+0.25}_{-0.22} M_\odot$ in
Table~\ref{tab:phys}.

We then add a flux constraint (as in Section~\ref{sec:flux}).  The
result is shown as blue contours in Figure~\ref{fig:bayes}.  As expected,
the effect is to sharply reduce the number of high-mass lenses.  For
example, the median host mass for
$(u_0>0)$ becomes $0.88^{+0.27}_{-0.22}\,M_\odot$ compared to 
$M_{\rm host}=1.14^{+0.10}_{-0.12} M_\odot$ in Table~\ref{tab:phys_flux}.  One may
compare the $1\,\sigma$ ranges of the two sets of distributions directly
in Tables~\ref{tab:phys_flux} and \ref{tab:bayes_flux_cond}.  Overall
the latter are two-to-four times broader, with peaks that are offset by
less $<1\,\sigma$.  That is, the result of the {\it Spitzer} parallax
measurement is to much more precisely locate the solution (despite the
absence of Galactic priors) within
the region expected in the absence of {\it Spitzer} data (but with Galactic
priors).  The main effect of the {\it Spitzer} data is to exclude
low mass lenses.  But these low-mass (high $\pi_\e$) lenses are already
significantly disfavored in the ground+Bayes analysis.

\begin{table}
\begin{center}
\caption{\textsc{Physical parameters from Bayesian analysis with flux constraint}}
\begin{tabular}{lcc}
\hline
\hline
\multicolumn{1}{c}{Parameter} &
\multicolumn{1}{c}{$u_0>0$} &
\multicolumn{1}{c}{$u_0<0$} \\
  $M_{\rm host}$ $[M_\odot]$     &$0.88_{-0.23}^{+0.27}$  &$0.79_{-0.27}^{+0.32}$  \\
  $M_{\rm planet}$ $[M_\oplus]$ &$5.35_{-1.38}^{+1.61}$  &$4.75_{-1.61}^{+1.93}$  \\
  $a_{\perp}$ [au]               &$3.60_{-0.69}^{+0.81}$  &$3.31_{-0.89}^{+1.00}$  \\
  ${\it D_L}$ [kpc]             &$2.82_{-0.73}^{+1.12}$  &$2.50_{-0.78}^{+1.22}$  \\
\hline
\end{tabular}
\label{tab:bayes_flux_cond}
\end{center}
\end{table}

\section{{Discussion}
\label{sec:discuss}}

KMT-2018-BLG-0029Lb has the lowest planet-host mass ratio $q=0.18\times 10^{-4}$
of any microlensing planet to date.  Although eight planets had previously
been discovered in the range of 0.5--1.0$\times 10^{-4}$, including seven
analyzed by \citet{ob171434} and one discovered subsequently \citep{ob180532},
none came even within a factor of two of the planet that we report here.
This discovery therefore proves that the previously discovered
pile-up of planets with Neptune-like planet-host mass ratios does
not result from a hard cut-off in the underlying distribution of planets.
However, it will require more than
a single detection to accurately probe the frequency of planets in
this sub-Neptune mass-ratio regime.  It is somewhat sobering that
after 16 years of microlensing planet detections there are only nine
with well measured mass ratios\footnote{Note that while OGLE-2017-BLG-0173L
\citep{ob170173} definitely has a mass ratio $q<1\times 10^{-4}$, it
is not included in this sample because it has two degenerate solutions
with substantially different $q$, and hence its mass ratio
cannot be regarded as ``well measured''.} $q\leq 1\times 10^{-4}$.
Hence, it is worthwhile to ask about the prospects for detecting more.

\subsection{{Prospects for Very Low $q$ Microlensing Planets}
\label{sec:vlqps}}

Of the nine such events, five were found 2005--2013 and four were found
2016--2018.  These two groups have strikingly different characteristics.
Four 
(OGLE-2005-BLG-390,
OGLE-2007-BLG-368,
MOA-2009-BLG-266, and
OGLE-2013-BLG-0341) from the first group revealed their planets via 
planetary caustics, and
only one (OGLE-2005-BLG-169) via central or resonant caustics.  By
contrast, all four from the second group revealed their planets 
via central or resonant caustics and all with impact parameters 
$u_0\la 0.05$.  Another telling difference is that follow-up observations
played a crucial or very important role in characterizing the planet
for four of the five in the first group\footnote{For the fifth,
OGLE-2013-BLG-0341L \citep{ob130341}, 
there were also very extensive follow-up observations,
which were important for characterizing the binary-star system
containing the host, but these did not play a major role in 
the characterization of the planet itself.}, while follow-up observations
did not play a significant role in characterizing any of the four planets
in the second group.  Finally, the overall rate of discovery approximately
doubled from the first to the second period.

The second period, 2016--2018, coincides with the full 
operation of KMTNet in its wide-field, 24/7 mode \citep{kmtk2c9,kmt2016}.
The original motivation for KMTNet was to find and characterize
low-mass planets without requiring follow-up observations \citep{eventfinder}.
All four planets from the second group were intensively observed by KMTNet, 
with the previous three all in high-cadence ($\Gamma=4\,{\rm hr}^{-1}$)
fields and KMT-2018-BLG-0029Lb in a $\Gamma=1\,{\rm hr}^{-1}$ field.
It should be noted that OGLE-2016-BLG-1195Lb was discovered and independently
characterized (i.e., without any KMTNet data) by OGLE and MOA
\citep{ob161195a}.  In this sense, it is similar to OGLE-2013-BLG-0341LBb,
which would have been discovered and characterized by OGLE and MOA data,
even without follow-up data.

The above summary generally confirms the suggestion of \citet{ob171434}
that the rate of low-mass planet discovery has in fact doubled in the
era of continuous wide-field surveys.  However, it also suggests that
this discovery mode (i.e., without substantial follow-up observations)
is ``missing'' many low-mass planets that were being discovered in
the previous period.  Apart from OGLE-2013-BLG-0341, which would have
been characterized without follow-up, three of the other four low-mass
planets from that period were all discovered in what would today be
considered ``outlying fields'', with Galactic coordinates $(l,b)$ of
OGLE-2005-BLG-169 $(0.67,-4.74)$,
OGLE-2007-BLG-368 $(-1.65,-3.69)$,
MOA-2009-BLG-266 $(-4.93,-3.58)$.  These regions are currently observed
by KMTNet at $\Gamma= (1,1,0.4)\,{\rm hr}^{-1}$.
Only OGLE-2005-BLG-390 $(2.34,-2.92)$ lies in what is now a 
high-cadence KMT field.

Moreover, the rate of discovery of microlensing events in these outlying
fields is much higher today than it was when these four planets were
discovered.  Hence, while there is no question that the pure-survey
mode has proved more efficient at finding low-mass planets, the
rate of discovery could be enhanced by aggressive follow-up observations.
See also Figure~8 from \citet{kb181292}.



\subsection{{Additional {\it Spitzer} Planet}
\label{sec:spitzplan}}


KMT-2018-BLG-0029Lb is the sixth published planet in the {\it Spitzer} 
statistical sample that is being accumulated to study the Galactic distribution
of the planets \citep{yee15,170event}.  
The previous five were\footnote{In addition, there
were two other {\it Spitzer} parallaxes for planets that are not
in the statistical sample,
OGLE-2016-BLG-1067Lb \citep{ob161067} and
OGLE-2018-BLG-0596Lb \citep{ob180596}.}
OGLE-2014-BLG-0124Lb \citep{ob140124},
OGLE-2015-BLG-0966Lb \citep{ob150966},
OGLE-2016-BLG-1190Lb \citep{ob161190},
OGLE-2016-BLG-1195Lb \citep{ob161195a,ob161195b}, and
OGLE-2017-BLG-1140Lb \citep{ob171140}.

While it is premature to derive statistical implications from this
sample, it is important to note that the planetary signature in the
KMT-2018-BLG-0029 light curve remained hidden in the real-time
photometry, although the pipeline re-reductions did yield strong hints
of a planet.  Nevertheless, TLC re-reductions were required for
a confident signal.  Hence, the history of this event provides
strong caution that careful review of all {\it Spitzer} microlensing
events, with TLC re-reductions in all cases that display possible hints of
planets, will be crucial for fully extracting information about
the Galactic distribution of planets from this sample.

\subsection{{High-Resolution Followup}
\label{sec:followup}}

As discussed in Section~\ref{sec:phys}, the blended light is 
consistent with being generated by the lens.  This identification
would be greatly strengthened if the blend (which is about 2 mag
brighter than the source in the $I$-band) were found to be
astrometically aligned with the position of fhe microlensed source
to the precision of high-resolution measurements.  These could
be carried out immediately using either ground-based adaptive optics
(AO) or with the {\it Hubble Space Telescope (HST)}.  Even if
such precise alignment were demonstrated, one would still have
to consider the  possibility that the blend was not the lens, but
rather either a star that was associated with the event 
(companion to lens or source), or even a random field star that was
not associated with the event.  These alternate possibilities
could be constrained by the observations themselves.  For example,
the blend's color and magnitude might be inconsistent with it lying 
in the bulge.  And the possibility that the blend was a companion
to the lens could be constrained by the microlensing signatures to which
such an object would give rise.  The possibility that the blend
is an ambient star could be estimated from the surface density of
stars of similar brightness together the astrometric precision of
the measurement.  It is premature to speculate on the analysis of
such future observations.  The main point is that these observations
should be taken relatively soon, before the lens and source substantially
separate, so that their measured separation reflects their separation
at the time of the event.

Even in the case that the relatively bright blend proves
to be displaced from the lens, these observations would still serve
as a first epoch to be compared to future high-resolution observations
when the lens and source have significantly separated.  If the lens
is sufficiently bright, its identification could be confirmed after
a relatively few years from, e.g., image distortion.  In the worst
case, the lens will not measurably add to the source flux, and so
could only be unambiguously identified when it had separated about
1.5 FWHM from the source.  This would occur
$\delta t = 3.2\,{\rm yr}(\lambda/1.1\,\mu{\rm m})(D/ 39{\rm m})^{-1}
(\mu/3.3\,\masyr)^{-1}$ after the event, where
$\lambda$ is the wavelength of observation and $D$ is the diameter
of the mirror.  Such observations would be feasible at first AO light
on any of the extremely large telescopes (ELTs) but would have to
wait until 2036 for, e.g., $1.6\,\mu{\rm m}$ observations on the Keck
10m telescope.

To assist in the interpretaion of such observations, we include
auxiliary files with the $(x,y,I)$ data for field stars on the same
system as the precision measurements for these quantities for the
microlensed source, namely $(x,y,I)=(151.96,149.30,22.02)$.

\acknowledgments

{\bfq We thank the anonymous referee for an especially valuable report
that helped greatly to clarify the issues presented here}.
Work by AG was supported by AST-1516842 from the US NSF
and by JPL grant 1500811.
AG received support from the European  Research  Council  under  the  European  Union’s Seventh Framework Programme (FP 7) ERC Grant Agreement n. [321035].
Work by C.H. was supported by the grant (2017R1A4A1015178) of the National Research Foundation of Korea.
This research has made use of the KMTNet system operated by the Korea
Astronomy and Space Science Institute (KASI) and the data were obtained at
three host sites of CTIO in Chile, SAAO in South Africa, and SSO in
Australia.
We are very grateful to the instrumentation and operations teams at CFHT who fixed several failures of MegaCam in the shortest time possible, allowing its return onto the telescope and these crucial observations. W.Z.and S.M. acknowledges support by the National Science Foundation of China (Grant No. 11821303 and 11761131004). MTP was supported by NASA grants NNX14AF63G and NNG16PJ32C, as well as the Thomas Jefferson Chair for Discovery and Space Exploration. This research uses data obtained through the Telescope Access Program (TAP), which has been funded by the National Astronomical Observatories of China, the Chinese Academy of Sciences, and the Special Fund for Astronomy from the Ministry of Finance.

\section{{Appendix: {\it Spitzer} Light-Curve Investigation}
\label{sec:syst}}

\begin{figure}
\centering
\includegraphics[width=90mm]{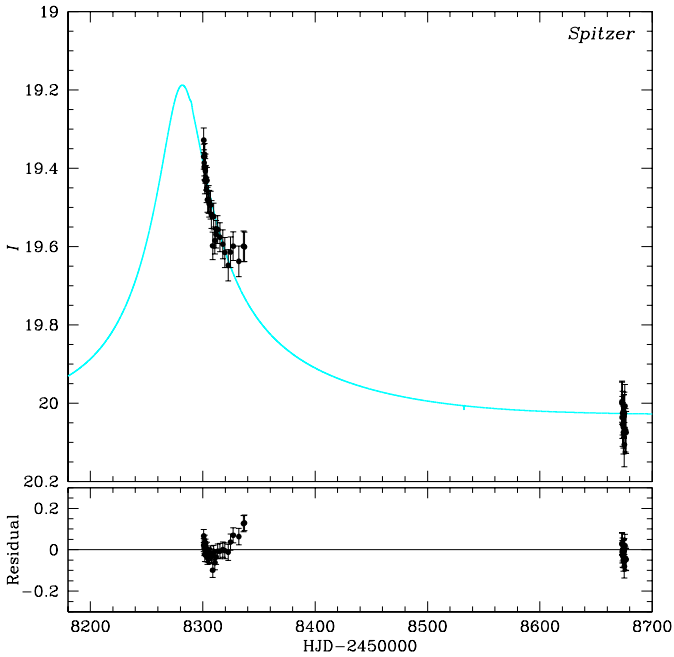}
\caption{Full {\it Spitzer} light curve, including all 2018 and
2019 data.  Compare to Figure~\ref{fig:lcspitz}, which shows the subset of
these data used in the analysis.  The truncated data set excludes the
second-half-2018 data, i.e., HJD$^\prime>8307$.  The full data set shown
here exhibits strong correlated residuals (high points at HJD$^\prime>8307$),
which then also induce high residuals in the first few points.  Note
that no such systematically high points are seen in Figure~\ref{fig:lcspitz}
because the model is freed from the necessity to try to fit the later
correlated high points.
}
\label{fig:lcfullspitz}
\end{figure}

The full {\it Spitzer} light curve (i.e., all-2018 plus 2019) exhibits
clear systematics, or more formally, residuals that are correlated
in time and with rms amplitude well above their photon noise.
This can be seen directly by comparing the full light curve 
(Figure~\ref{fig:lcfullspitz}) to the one analyzed in the main
body of the paper (Figure~\ref{fig:lcspitz}).  In addition to the
clear correlated residuals in the latter, it also has an error 
renormalization factor (relative to the photon-noise-based pipeline
errors) of 2.30 compared to 1.17\footnote{Note that this is just barely
above the $1\,\sigma$ range $1\pm (2 N_{\rm dof})^{-1/2}\rightarrow
1\pm 0.14$ for uncorrelated, purely Gaussian statistics with $N_{\rm dof}=26$
degrees of freedom.} when the second-half-2018 data are
removed.

\begin{figure}
\centering
\includegraphics[width=90mm]{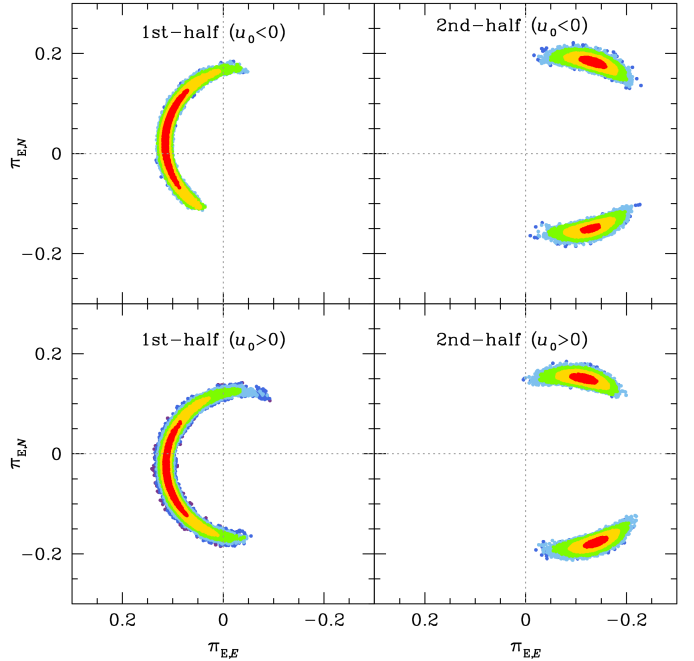}
\caption{{\it Spitzer}-``only'' parallax contours for two subsets
of the data, either ``1st-half''  or ``2nd-half'' 
of 2018 {\it Spitzer} data together with all (i.e., one week) of 
2019 (``baseline'') {\it Spitzer} data.  The contours 
for $u_{0,\oplus}< 0$ (upper) and $u_{0,\oplus}> 0$ (lower) are similar,
but the contours based on ``1st-half''  (left) and ``2nd-half'' (right)
are very different.  The tension between these two halves is a reflection
of the correlated photometry errors as seen in Figure~\ref{fig:lcfullspitz}
}
\label{fig:1st2nd}
\end{figure}

A second way to view the impact of these correlated errors is to compare
the {\it Spitzer}-``only'' solution derived from combining
first-half-2018 with 2019 data to the one derived from combining
second-half-2018 with 2019 data.  See Figure~\ref{fig:1st2nd}.
While the upper $(u_0>0)$ and lower $(u_0<0)$ pairs of panels are
similar, the left (first-half) and right (second-half)
pairs of panels are radically different.  They have completely
different morphologies, and the contours themselves only overlap
at the $3\,\sigma$ level.

Yet a third way to view the impact of these correlated errors is to
``predict'' the ``baseline'' {\it Spitzer} flux, 
$f_{\rm base}\equiv f_s + f_b$ from the full 2018 data set and then
compare this with the measured $f_{\rm base}$ from 2019 data.  this
yields $3.0\pm 0.1$ versus $2.46\pm 0.03$.

While these are just different ``viewing angles'' of the same
effects in the data, we present all three because they open different
paths to trying to establish their origin.  Any attempt
to identify a physical cause for these effects must begin with a
physical understanding of the measurement process together with the
specific physical conditions of the measurment.

The data stream consists of six dithered exposures at each epoch,
each of which yields a matrix of photo-electron counts from the detector.  
In contrast to optical CCDs, the PRF of the detector is highly non-uniform over
the pixel surface, which means that the quantitative interpretation
of the pixel counts in terms of incident photons requires relatively
precise knowledge of the stellar positions in the frame of the detector
matrix.  This applies both to the target star as well as any other
stars whose light profile (PSF) signficantly overlaps that of the
target.  We note that this would not be true if 1) one were interested
in only relative photometry and 2) the detector position and orientation
returned to the same sky position and orientation 
(or set of six sky positions and orientations) at each epoch.  In that
case, one could use a variant of DIA.  However, neither
condition applies to {\it Spitzer} microlensing observations. 
Most importantly, the observations typicaly span four to six weeks,
during which the detector rotates by several degrees.  In addition,
one must actually know the target position in order to translate
total photon counts into a reliable estimate of incident photons,
which in turn is required to apply the $VIL$ (or $IHL$) color-color 
relation.
This latter problem is usually solved with adequate precision.
However, the impossibility of DIA, together with the constraints
imposed by crowded fields, is what led to the development of a new
PRF photometry algorithm \citep{170event}.

This algorithm operates with several variants.  For example, if the
source is relatively bright at all epochs, then its position can be
determined on an image-by-image basis.  If it is bright at some epochs
and not others, then the first group can be used to determine the source
position relative to a grid of field stars, with this position then
applied to the second group.  If the source position cannot be determined
at all from the {\it Spitzer} data (e.g., because the event is well
past peak by the time the observations begin), then it can be found
near peak from DIA of optical data relative to a grid of optical field stars.
Then this optical grid can be cross-matched to {\it Spitzer} field stars,
which leads to a prediction of the source position relative to the detector
matrix.  In general, one of more of these procedures works quite well
for the great majority of {\it Spitzer} microlensing events that are subjected
to TLC analysis.

However, for KMT-2018-BLG-0029, the conditions were especially challenging.
First, the source flux (determined from the color-color relation)
$f_{s,Spitzer} = 0.58$ is quite small relative to that of three blends that
lie within about 2 pixels, i.e., 40, 35, and 29.  Second these bright
blends overlap each other (and possibly other unresolved stars), and
hence it is impossible to reliably determine their positions even
from the higher-resolution ground-based data.  (By contrast, although
the source is much fainter than the neighboring blends, its position 
can be derived from ground-based DIA because it varies strongly.)

One initially plausible conjecture for the origin of the correlated errors
would be that
the photometry is more reliable when the source is brighter simply
because its position is better determined on an epoch-by-epoch basis,
and that the poorly known positions of the blends increasingly corrupt
the measurements when the source is fainter.  This conjecture would
lead to the following ``triage sequence'' of confidence in the data:
first-half-2018, second-half-2018, 2019, i.e., by decreasing brightness.
Moreover, tests show that the target centroid can be constrained for
almost all of the 2018 epochs based on {\it Spitzer} data alone,
typically to within $\sim 0.1$ and $\sim 0.2$ pixels per epoch, 
for the first and second halves, respectively, but cannot be constrained
at all for 2019.  This line of reasoning would possibly lead to accepting all
the 2018 data and rejecting the 2019 data on the grounds that the
2019 data were ``most affected by systematics''.  

We considered this approach but rejected it for reasons that are given
in the next paragraph.  Our main reason for recounting it in some detail
is to convince the reader of its superficial plausibility and also of the danger
of ``explaining'' evident correlated residuals by ``phyiscal'' arguments that
are not rooted in the real physical conditions.  We note that the
interested reader can see the result of applying this approach by
accessing the version of this paper that was prepared prior to the
2019 {\it Spitzer} microlensing season, i.e., when only 2018 data were
available (arXiv:1906.11183).  In fact, the final results derived from 
this 2018-only analysis do not differ dramatically from those presented
in the body of this paper, although some of the intermediate steps look
quite different.

The first point to note is that there is an immediate warning flag
regarding this approach: the 2018-only light curve looks much worse
(arXiv:1906.11183) than the first-half-2018-plus-2019 light curve and,
correponding to this, has a much higher error-renormalization factor.
This already suggests (although it hardly proves) that the real problems
are concentrated in the second-half-2018 data.  However, more fundamentally,
the logic on which the conjecture is based does not hold up.  The
centroid position can be determined to better than 0.1 pixels by transforming
from the optical frame, so the fact that this centroid can be determined
to 0.2 pixels from the second-half-2018 data has no practical implication
for the photometry.  And in particular, the same correlations between the 
residuals remain for the second-half-2018 data whether the position is derived
from {\it Spitzer} images alone or by transformation from the optical
frame.

Another path toward understanding this issue, which proves
to be more self-consistent, is to examine the photometry of the
three bright blends as a function of time.  In all three cases, the
mean value and scatter are very similar when the first-half-2018
and 2019 data are compared.  These are
[($39.62\pm 0.13,0.51$) versus ($39.76\pm 0.16,0.61$)],
[($34.74\pm 0.49,1.91$) versus ($35.22\pm 0.44,1.64$)], and
[($28.90\pm 0.69,2.68$) versus ($28.76\pm 0.55,2.05$)] 
for the first, second, and third blend, respectively.  
That is, the means differ by 
$0.68\,\sigma$, 
$0.73\,\sigma$, and
$0.16\,\sigma$, respectively.
By contrast, both the first and third blend
display strong ``features'' during $8310<{\rm HJD}^\prime<8338$.
For the first blend, these data have a mean of $40.69\pm 0.20$, i.e., 
$5\,\sigma$
higher than predicted by the combined first-half-2018 and 2019 data:
$39.68\pm 0.10$
For the third blend, these data have similar mean but a scatter (5.38)
that is well over twice the values of the other two periods.
This is strong empirical evidence that the first-half-2018 and 2019
data are rooted in a comparable physical basis, but the second-half-2018
data are not.  Given that the field rotation, in combination with the
severe crowding from several bright blends, provide a plausible physical
explanation for these differences, we conclude that first-half-2018 and 2019
data can be analyzed as a single data set, but the second-half-2018 data
must be excluded from the analysis.

\end{document}